\documentclass[conference]{IEEEtran}
\IEEEoverridecommandlockouts
\usepackage{cite}
\usepackage{amsmath,amssymb,amsfonts}
\usepackage{algorithm}
\usepackage[noend]{algpseudocode} 
\algrenewcommand\algorithmicindent{0.8em}
\algrenewcommand\alglinenumber[1]{\scriptsize #1}
\algrenewcommand\textproc{\textsc}   
\usepackage{graphicx}
\usepackage{textcomp}
\usepackage{xcolor}
\usepackage{listings}
\usepackage{xcolor}  %
\usepackage{subcaption}
\usepackage{etoolbox}
\newcommand{\Indent}{\hspace*{1em}}
\newcommand{\EndIndent}{}
\usepackage[margin=1in]{geometry}
\usepackage{hyperref}
\usepackage{comment}
\usepackage{subcaption}
\usepackage[margin=1in]{geometry}
\usepackage{booktabs} 
\usepackage{amsmath}  
\usepackage{tabularx} 
\usepackage{makecell} 
\usepackage{minted}
\usepackage{xcolor}
\usepackage{caption}
\usepackage{adjustbox}
\usepackage{array}
\usepackage{url}

\def\BibTeX{{\rm B\kern-.05em{\sc i\kern-.025em b}\kern-.08em
    T\kern-.1667em\lower.7ex\hbox{E}\kern-.125emX}}

\author{
\IEEEauthorblockN{Adiba Masud, Nicholas Foley, Pragathi Durga Rajarajan, and Palden Lama} 

\IEEEauthorblockA{Department of Computer Science\\
The University of Texas at San Antonio, San Antonio, Texas, USA\\
\{adiba.masud, nicholas.foley, durga.rajarajan\}@my.utsa.edu, palden.lama@utsa.edu}

}

\begin{document}
\title{Where to Split? A Pareto-Front Analysis of DNN Partitioning for Edge Inference}
\maketitle

\begin{abstract}
The deployment of deep neural networks
(DNNs) on resource-constrained edge devices is frequently
hindered by their significant computational and memory
requirements. While partitioning and distributing a DNN across multiple devices is a well-established strategy to mitigate this challenge, prior research has largely focused on single-objective optimization, such as minimizing latency or maximizing throughput. This paper challenges
that view by reframing DNN partitioning as a multi-objective
optimization problem. We argue that in real-world
scenarios, a complex trade-off between latency and
throughput exists, which is further complicated by network
variability. To address this, we introduce ParetoPipe, an
open-source framework that leverages Pareto front analysis
to systematically identify optimal partitioning strategies
that balance these competing objectives.

Our contributions are threefold: we benchmark pipeline partitioned inference on a heterogeneous testbed of Raspberry
Pis and a GPU-equipped edge server; we identify
Pareto-optimal points to analyze the latency-throughput
trade-off under varying network conditions; and we release
a flexible, open-source framework to facilitate distributed
inference and benchmarking. This toolchain features dual
communication backends, PyTorch RPC and a custom
lightweight implementation, to minimize overhead and
support broad experimentation.

\end{abstract}


\section{Introduction}
The deployment of deep neural networks (DNNs) at the network edge is crucial for enabling real-time, high-throughput decision-making in applications from smart home automation to industrial IoT \cite{liu2019intelligent,dai2019industrial,sodhro2019ai,liu2021intelligent}. By localizing processing on ARM-based devices or compact GPU-equipped platforms, this paradigm reduces latency and reliance on the cloud. However, it also faces a critical bottleneck: the computational and memory demands of state-of-the-art DNNs overwhelm the capabilities of such resource-constrained hardware~\cite{zeng2021coedge}. To mitigate this, pipeline partitioning, distributing a DNN across a network of devices, has emerged as a promising strategy for balancing load and improving performance.

While prior works have advanced distributed inference by partitioning models to either minimize inference latency or maximize throughput, this has been largely treated as a single-objective problem \cite{feltin2023dnn,li2024distributed,PipeEdge,wang2023model}. We contend that in realistic deployments, no single "best" solution exists. Instead, practitioners must navigate a complex trade-off space, most notably between latency and throughput, where performance is further complicated by unpredictable network delays and bandwidth constraints. Therefore, we reframe DNN partitioning as a multi-objective optimization problem. To this end, we introduce \emph{ParetoPipe}, an extensible open-source framework designed to systematically identify the entire Pareto frontier—the set of all partitioning strategies where one objective (e.g., inference latency) cannot be improved without degrading another (e.g., throughput). ParetoPipe enables a holistic analysis on heterogeneous edge hardware, complete with multiple communication backends to investigate the impact of runtime overhead.

Our key contributions are:
\begin{itemize}
    \item The design and implementation of ParetoPipe, an extensible open-source framework for orchestrating, benchmarking, and visualizing the latency-throughput Pareto frontier for distributed DNNs. It features dual communication backends (PyTorch RPC and a custom lightweight implementation) to enable fine-grained analysis of runtime overhead.
    \item A comprehensive analysis of the Pareto-optimal frontiers for pipeline partitioning, generated using ParetoPipe across multiple DNN architectures on a heterogeneous testbed of Raspberry Pis and GPU-equipped edge server. This reveals the precise trade-offs that practitioners must navigate between latency and throughput.
    \item A empirical study of the Pareto frontier's sensitivity to network delay and bandwidth, yielding practical guidelines for selecting robust partitioning strategies that remain optimal under the variable conditions of real-world edge deployments.
   
\end{itemize}

The rest of the paper is organized as follows. Section 2 reviews the related works in edge inference. Section 3 details our system design and implementation. Section 4 describes the experimental setup. Section 5 presents evaluation results. Finally, Section 6 concludes the paper.

\section{Background and Related Work}
\subsection{Distributed DNN Inference:} 
When a single device has insufficient resources to store or efficiently run a DNN, the model must be partitioned across multiple devices. This can be achieved through as tensor parallelism, or pipeline parallelism. Tensor parallelism \cite{hu2024edge,qin2024disco,sun2024partitioned} divides individual layers of a model across several devices. While this enables execution of large models, it is communication-intensive and demands high-bandwidth interconnects, making it impractical for most edge network environments. In contrast, pipeline parallelism \cite{maruf2024optimizing,jafari2024pisel} splits the model sequentially, assigning consecutive layers to different devices to form a pipeline. The output of one device is passed as the input to the next. This method aligns well with the typical communication patterns and bandwidth limitations of edge networks. Given these advantages, this work focuses exclusively on the pipeline paradigm to analyze its performance for edge inference.

\subsection{Model Partitioning for Edge Devices:}
Within the pipeline parallelism paradigm, the primary research challenge is to identify the optimal points at which to split the model. Several pioneering frameworks have been developed to address this partitioning problem. Wang et al. \cite{wang2023model} proposed a partitioning framework that leverages computation and network profiles to optimize latency-aware distributed DNN inference across heterogeneous edge devices. Li et al. \cite{li2024distributed} used a multi-task learning-based asynchronous advantage actor-critic (A3C) algorithm to determine how to split a DNN model into multiple blocks for collaborative execution across edge servers and IoT devices. This technique was designed to find a partitioning policy that reduces the overall DNN inference latency. Feltin et al. \cite{feltin2023dnn} present a method to accelerate DNN inference throughput by partitioning the network across multiple edge devices using a branch and bound algorithm. Hu et al. \cite{PipeEdge} introduce EdgePipe, a framework that uses a dynamic programming algorithm to optimally partition and pipeline large model inference across heterogeneous edge devices, improving inference throughput. 

However, prior works optimize for either latency or throughput, leaving the fundamental trade-off between these metrics in distributed, heterogeneous environments unexplored. Our work fills this gap, but instead of focusing on a fast optimization algorithm, we conduct a comprehensive Pareto-front analysis. The goal is to thoroughly map the frontier of optimal latency-throughput solutions under various network conditions, thereby establishing a vital performance baseline and providing a toolchain to facilitate future research.

\section{System Design and Implementation}

\subsection{ParetoPipe System Architecture:}

\begin{figure}
    \centering
    \includegraphics[width=0.4\textwidth]{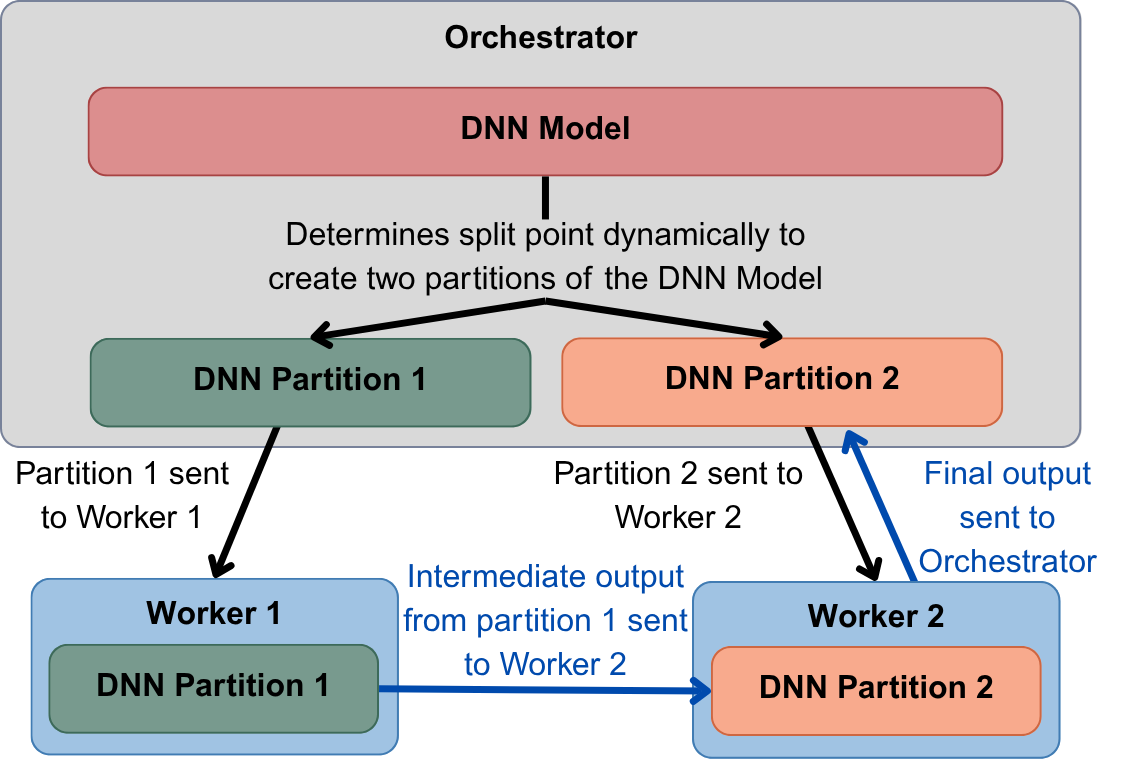}
    \caption{The ParetoPipe system architecture. The orchestrator partitions a DNN model across a set of heterogeneous worker nodes, creating a multi-stage pipeline. Each worker processes its assigned stage and forwards intermediate results to the next device in the pipeline.}
    \label{fig:system_architecture}
\end{figure}

ParetoPipe leverages pipeline parallelism to enable efficient execution across heterogeneous edge devices. As shown in Fig.~\ref{fig:system_architecture}, the key components include an \emph{orchestrator} and multiple \emph{workers}. The orchestrator, hosted on an edge server, is responsible for managing the distributed environment, deploying partitioned model segments to the respective edge devices, and managing inference flow. The worker processes run on each edge device and the edge server to execute inference on a designated partition of the DNN model. The orchestrator begins the distributed workflow by splitting a DNN model into two segments: the first is dispatched to worker 1 and the second to worker 2. The worker 1 processes each batch of input data through its assigned DNN model partition and sends the intermediate output to worker 2, which completes the remaining computation and returns the final prediction to the orchestrator. 


\subsection{Communication Backend}

ParetoPipe features dual communication backend using PyTorch RPC and a lightweight implementation. The PyTorch RPC approach leverages PyTorch’s built-in distributed communication framework, offering a high-level abstraction for model parallelism and inter-device communication. In contrast, our custom implementation handles distributed communication using TCP sockets, bypassing PyTorch’s RPC layer to minimize overhead and provide finer control over the execution flow.

\subsection{Algorithmic Framework of ParetoPipe}
To complement the system architecture in Fig.\ref{fig:system_architecture}, we summarize the proposed approach in a formal algorithmic framework. Algorithm \ref{alg:paretopipe} specifies the orchestrator logic, which dynamically determines split points, deploys model partitions, and coordinates distributed inference across heterogeneous workers. Algorithm \ref{alg:paretopipe} also specifies the worker logic loop, describing how each edge device loads its assigned partition, performs local inference, and exchanges intermediate or final outputs.


\begin{algorithm}[t]
\footnotesize
\caption{ParetoPipe Distributed Inference}
\label{alg:paretopipe}
\begin{algorithmic}[1]
\State Orchestrator initializes connections with Worker\_1 and Worker\_2
\State Orchestrator selects partition index \(p\) and signals workers:
    \Indent
      \State Layers \(L_{1..p}\) \(\rightarrow\) Worker\_1
      \State Layers \(L_{p+1..N}\) \(\rightarrow\) Worker\_2
    \EndIndent
\For{each input sample (or batch) \(x\)}
    \State Worker\_1: forward\((L_{1..p}, x)\), send activation \(A_p\) to Worker\_2
    \State Worker\_2: receive \(A_p\), forward\((L_{p+1..N}, A_p)\) to produce \(\hat y\)
    \State Worker\_2 \(\to\) Orchestrator: return \(\hat y\)
    \State Orchestrator collects and delivers \(\hat y\)
\EndFor
\State Orchestrator signals Workers to terminate
\end{algorithmic}
\end{algorithm}

\section{Experimental Setup}

\subsection{Hardware and Software:}

Our experimental testbed comprises two Raspberry Pi 4B devices with 4 core CPU and 4GB of memory, serving as edge nodes and an edge server equipped with an NVIDIA GeForce RTX 4090 GPU, 32 core CPU, and 124GB of memory. The edge server performs a dual role, acting as both the central orchestrator and a high-performance worker to leverage its GPU for accelerated DNN inference. All nodes communicate over a wired Ethernet LAN, with baseline network latencies (round-trip time) of 0.201 ms between the two Raspberry Pis and 0.383 ms between each Pi and GPU machine. To emulate realistic network conditions, we employed the Linux traffic control (tc) utility to impose network latency and bandwidth constraints during selected experiments. ParetoPipe was implemented using the PyTorch library and is publicly available on GitHub \cite{distributed-inference}.

\subsection{DNN Models}
\begin{table}[]
    \centering
    \begin{tabular}{c|c|c|c|c}
        \toprule
        \textbf{Model} & \textbf{\# Parameters}& \textbf{\# Blocks} & \textbf{Size} & \textbf{Accuracy}\\
         & &  & \textbf{(MB)} & \textbf{(\%)}\\
        \midrule
        MobileNetV2 & 2,236,682 & 21 & 8.8 & 90.1 \\
        ResNet18 & 11,689,512 & 14 & 43 & 90.0 \\ 
        InceptionV3 & 24,371,444 & 22 & 97 & 90.1 \\
        ResNet50 & 25,557,032 & 22 & 91 & 87.2 \\
        AlexNet & 61,100,840 & 21 & 234 & 79.6 \\ 
        VGG16 & 138,357,544 & 39 & 528 & 84.8 \\ 
        \bottomrule
    \end{tabular}
    \caption{DNN models used in benchmarking. }
    \label{tab:model_info}
\end{table}

To evaluate our distributed DNN framework, we selected a diverse set of convolutional neural network (CNN) architectures: MobileNetV2~\cite{mobilenetv2_paper}, ResNet18~\cite{resnets_paper}, InceptionV3~\cite{inceptionv3_paper}, ResNet50~\cite{resnets_paper}, AlexNet~\cite{alexnet_paper}, and VGG-16~\cite{vgg16_paper}. These models span a broad spectrum of computational complexity, layer diversity, and architectural depth, making them useful for evaluating our distributed inference frameworks under varying conditions. 

Table~\ref{tab:model_info} shows some key characteristics of the models. Here, a block refers to a group of layers that collectively perform a specific function and are often reused across architectures. For instance, many models, including AlexNet and the ResNet family, make use of convolutional blocks, while InceptionV3 utilizes multiple inception blocks.

We used the standard pretrained weights provided by PyTorch for each model (trained on ImageNet), and subsequently fine-tuned on the CIFAR-10~\cite{cifar10} dataset to improve model accuracy and adapt the models to CIFAR-10's input size (32x32) and number of output classes.


\begin{figure}[t]
    \centering

    \begin{subfigure}[t]{\linewidth}
        \centering
        \includegraphics[width=0.8\textwidth]{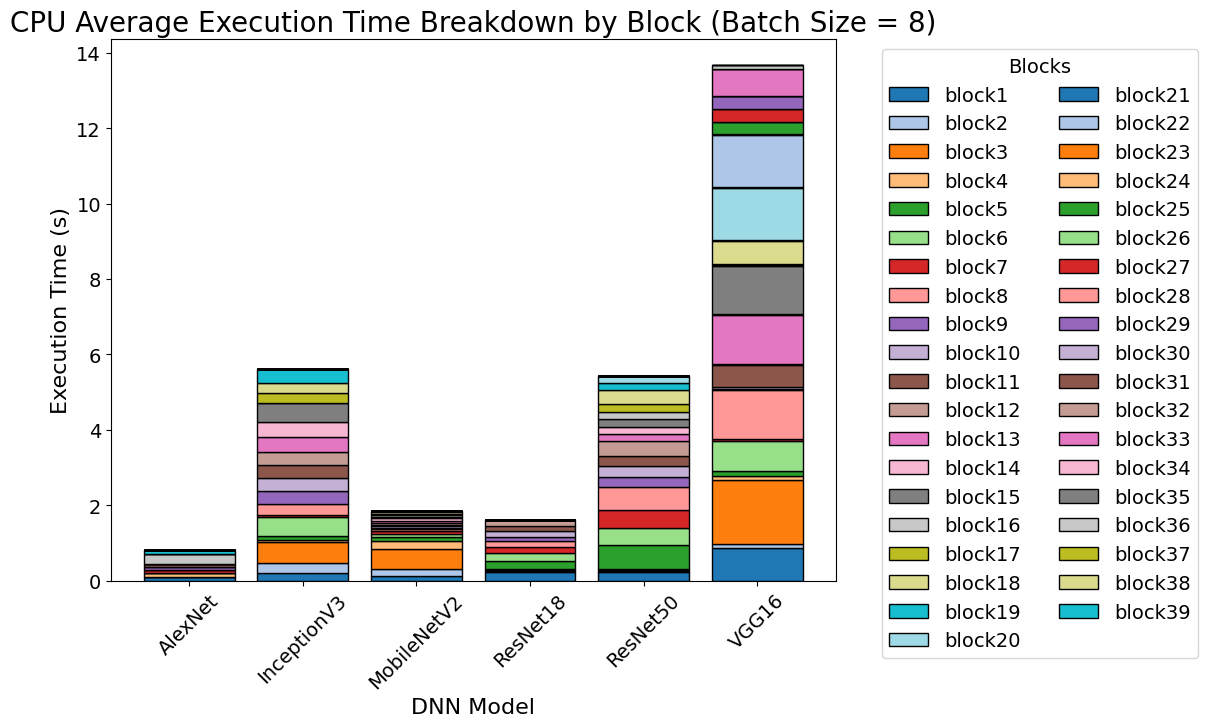}
        \caption{Block-wise execution times on CPU}
        \label{fig:cpu_blockwise_profiling_results_graph}
    \end{subfigure}

    \vspace{1em} 

    \begin{subfigure}[t]{\linewidth}
        \centering
        \includegraphics[width=0.8\textwidth]{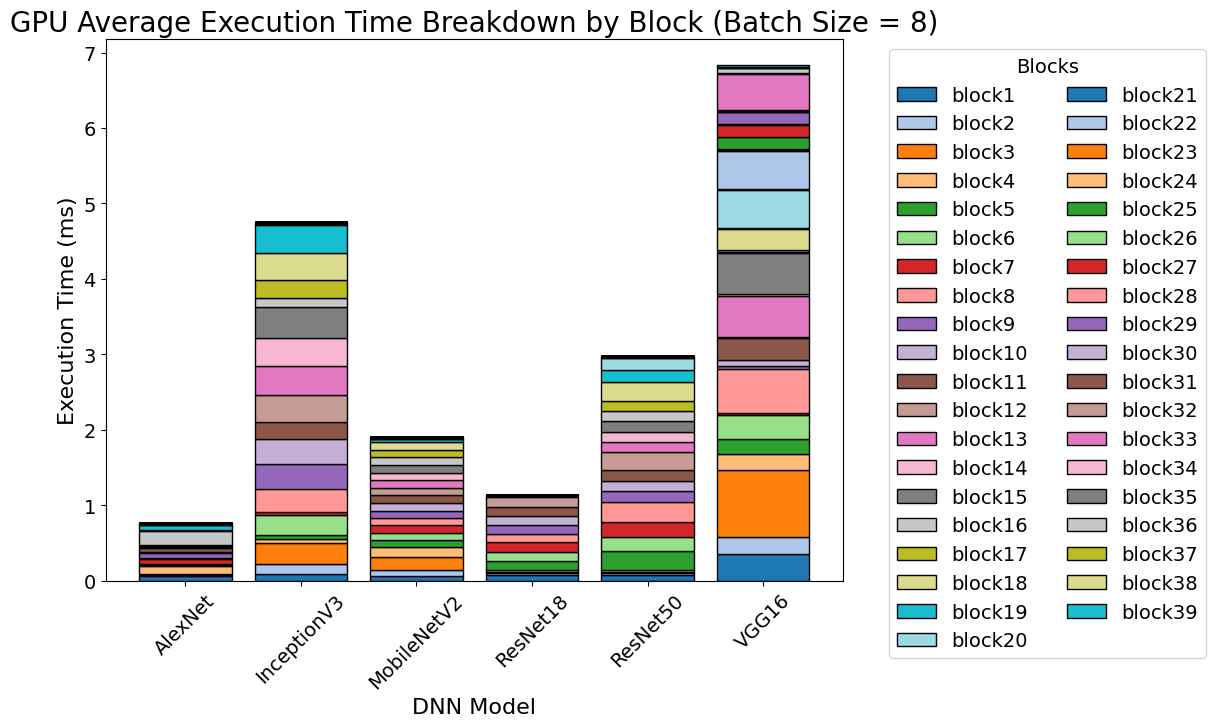}
        \caption{Block-wise execution times on GPU}
        \label{fig:gpu_blockwise_profiling_results_graph}
    \end{subfigure}

    \caption{Block-wise profiling results of DNN models on Raspberry Pi (CPU) and edge server (GPU). } 
    \label{fig:blockwise_profiling_combined}
\end{figure}

\subsection{Benchmarking Methodology:}
\label{sec:methodology}
Our primary performance metrics used for benchmarking include inference throughput (images/second) and end-to-end latency per batch. We also monitored CPU utilization, network load, and memory usage on both orchestrator and worker devices. All system metrics were collected using the psutil Python library.

Our evaluation methodology involved partitioning DNN models at the block boundaries rather than by individual layers. For each model, we exhaustively tested every valid partition point. To ensure statistically reliable measurements, each configuration was executed five times, and the results were averaged. We then performed a Pareto analysis on the collected performance (throughput and latency) to identify the optimal set of partition points that balance these competing objectives. 

\subsection{Block-wise Profiling of DNN Models:}
\label{sec:profiling}
We profiled the execution times of each block of the six models we tested on on a Raspberry Pi 4B's 4 core CPU and our Lambda server's NVIDIA GeForce RTX 4090 GPU. The profiling results in Fig.~\ref{fig:blockwise_profiling_combined} show that the execution times vary across blocks within the same model, highlighting that not all blocks are equal in computational cost. 

\begin{figure}[tbp]
  \centering
  \begin{subfigure}[b]{0.49\linewidth}
    \includegraphics[width=\linewidth]{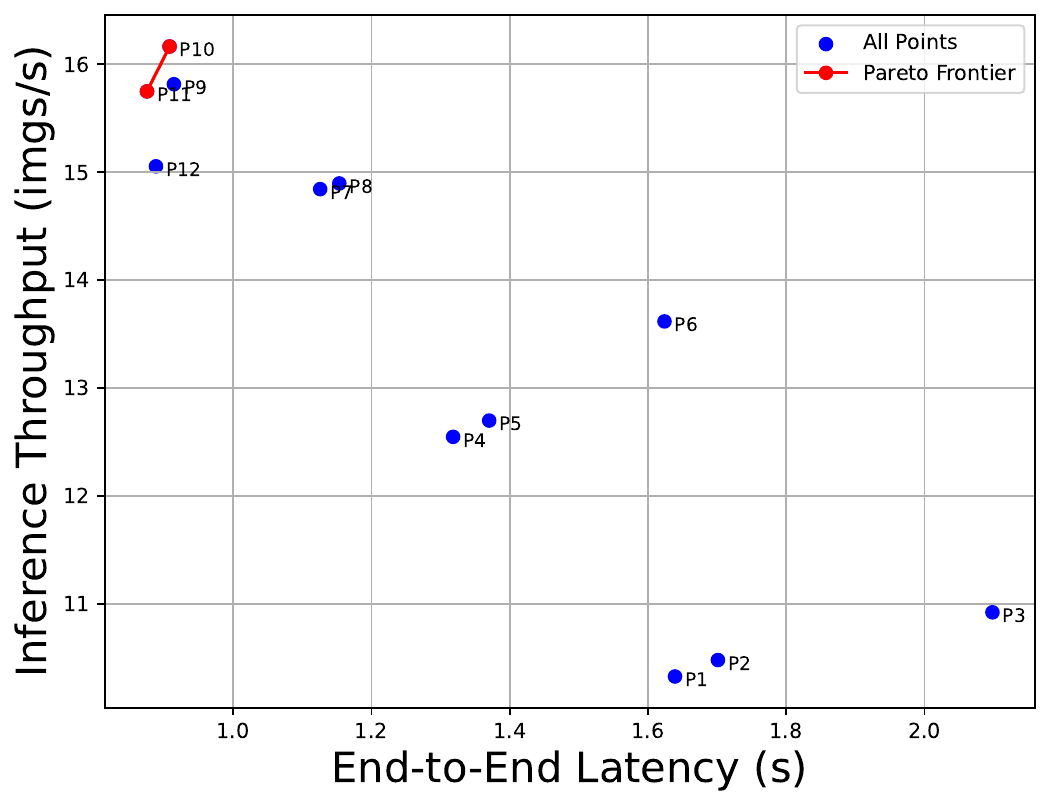}
    \caption{AlexNet}
  \end{subfigure}
  \begin{subfigure}[b]{0.49\linewidth}
    \includegraphics[width=\linewidth]{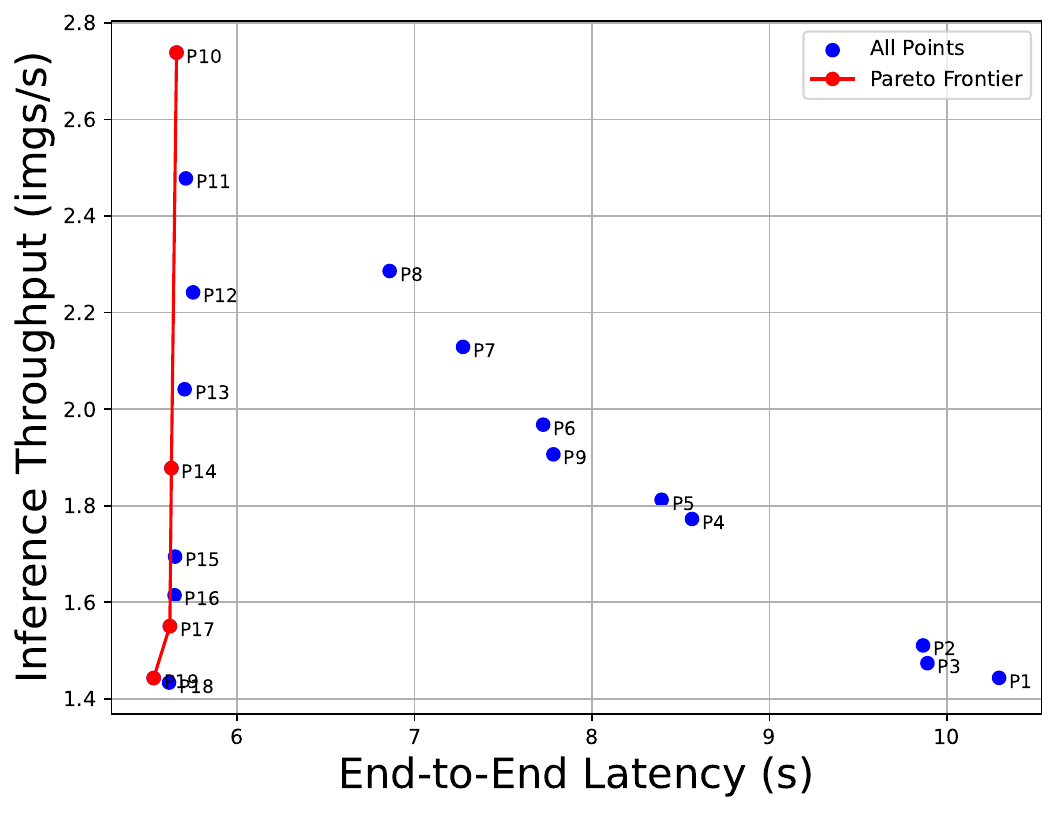}
    \caption{InceptionV3}
  \end{subfigure}
  \begin{subfigure}[b]{0.49\linewidth}
    \includegraphics[width=\linewidth]{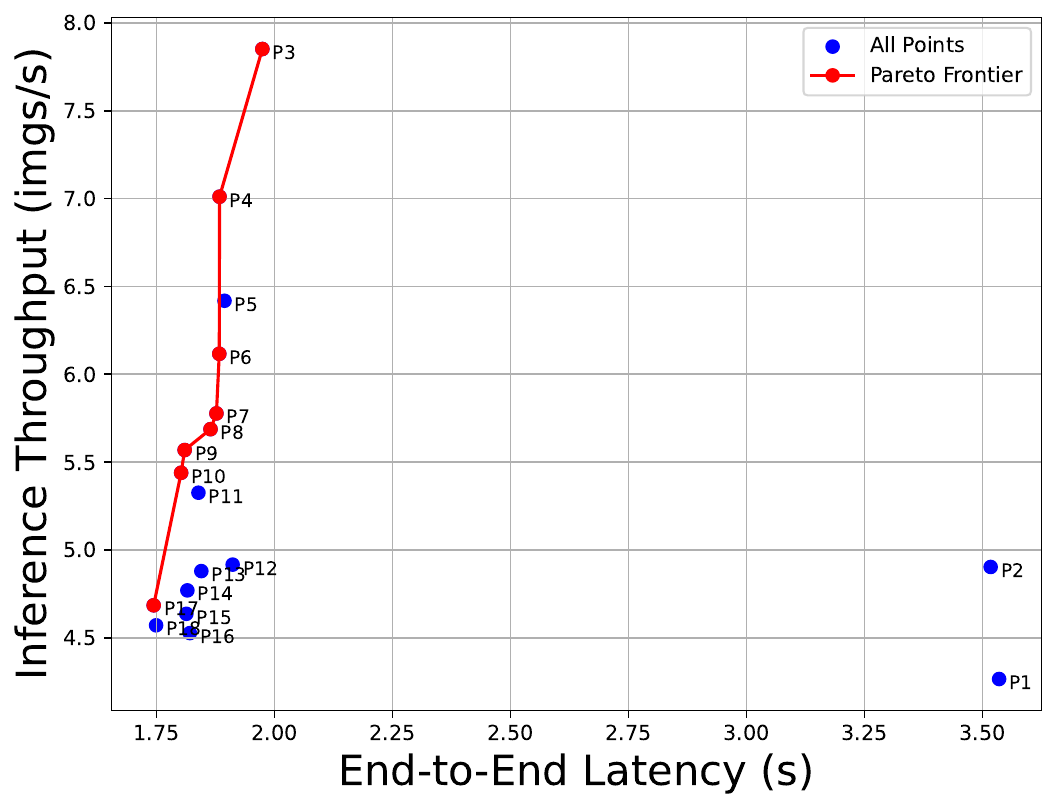}
    \caption{MobileNetV2}
  \end{subfigure}
  \begin{subfigure}[b]{0.49\linewidth}
    \includegraphics[width=\linewidth]{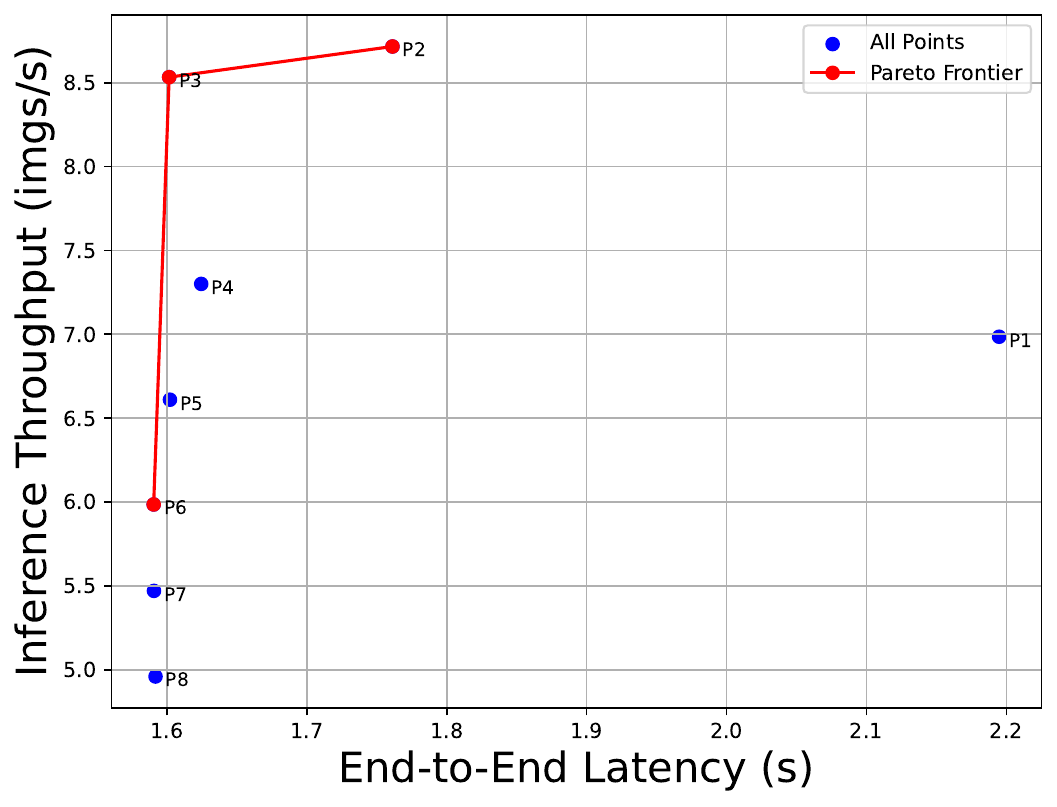}
    \caption{ResNet18}
  \end{subfigure}
  \begin{subfigure}[b]{0.49\linewidth}
    \includegraphics[width=\linewidth]{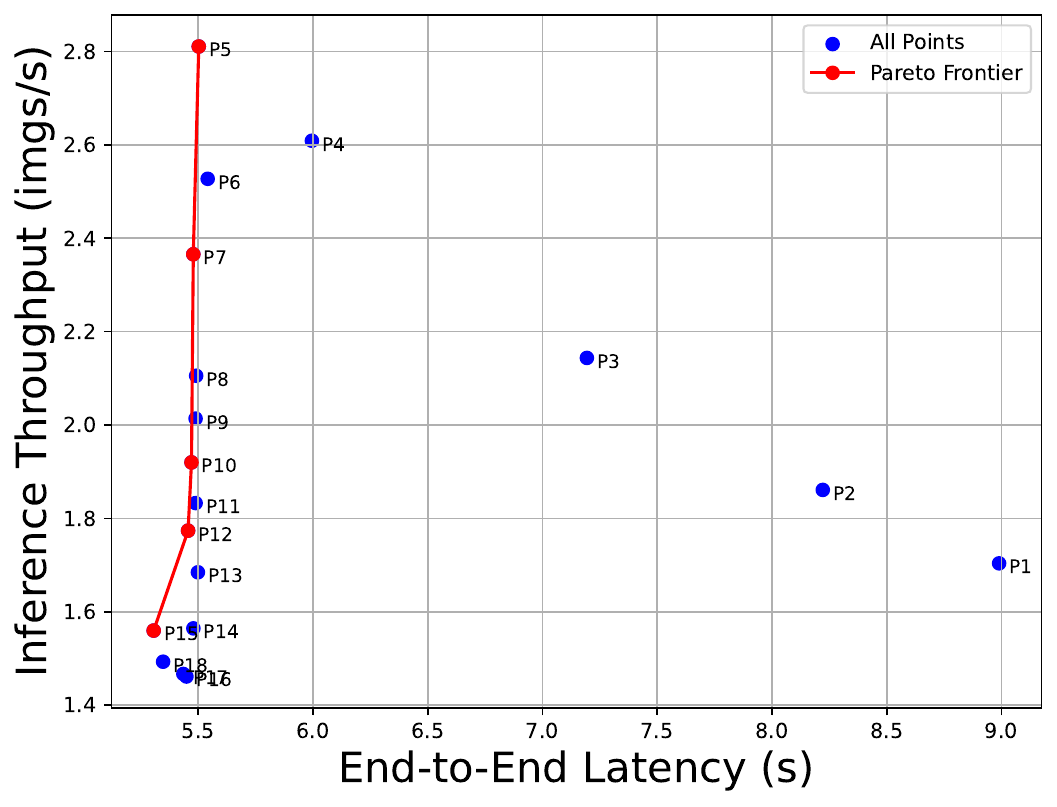}
    \caption{ResNet50}
  \end{subfigure}
  \begin{subfigure}[b]{0.49\linewidth}
    \includegraphics[width=\linewidth]{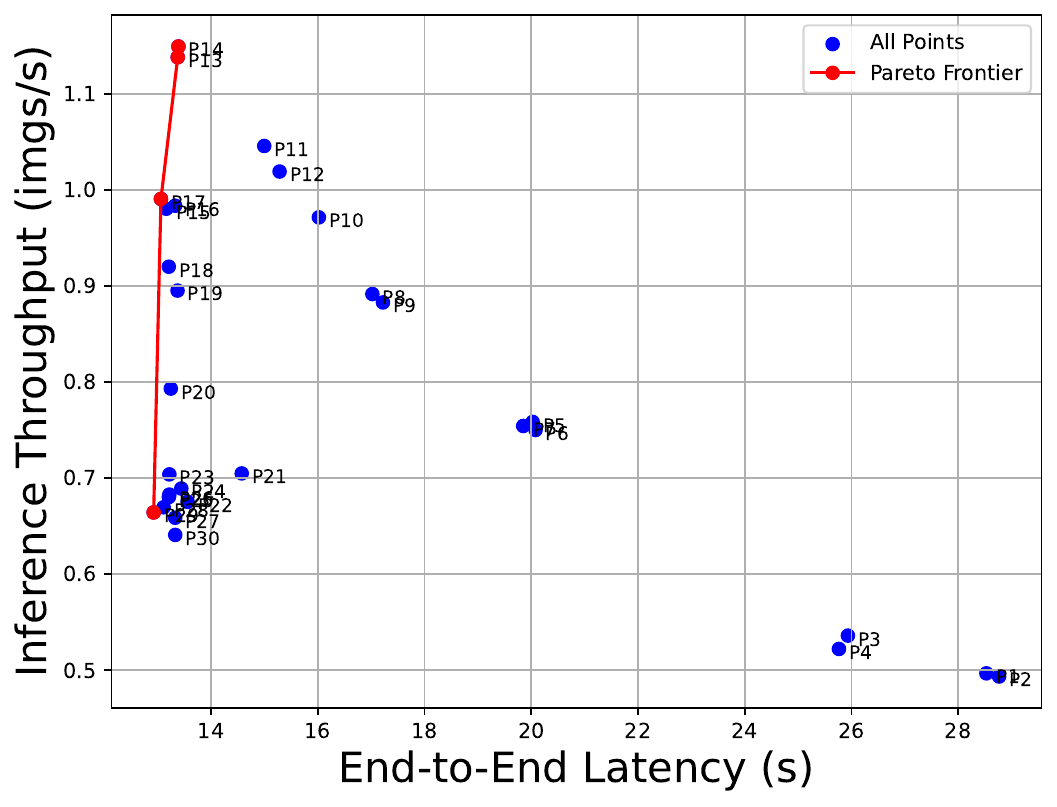}
    \caption{VGG16 }
  \end{subfigure}
  \caption{Pareto Frontiers for DNN Partitioning on a Pi-to-Pi configuration.}
  \label{fig:pi-pi}
\end{figure}

\begin{figure}[tbp]
  \centering
  \begin{subfigure}[b]{0.49\linewidth}
    \includegraphics[width=\linewidth]{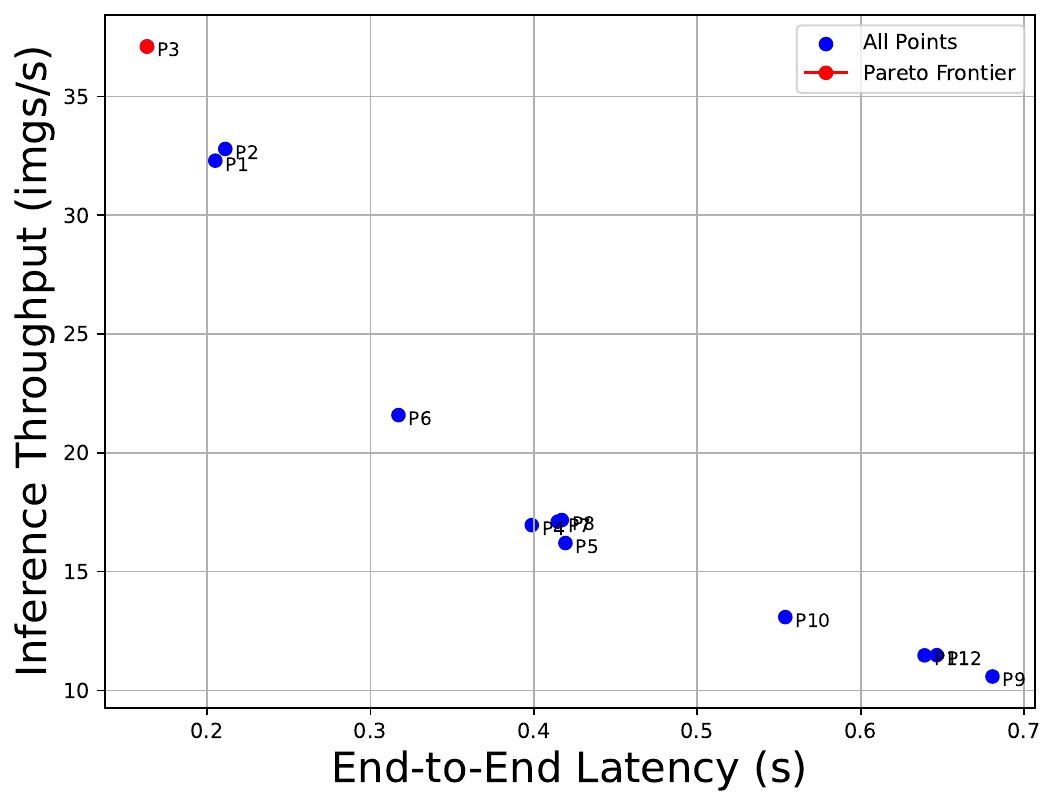}
    \caption{AlexNet}
  \end{subfigure}
  \hfill
  \begin{subfigure}[b]{0.49\linewidth}
    \includegraphics[width=\linewidth]{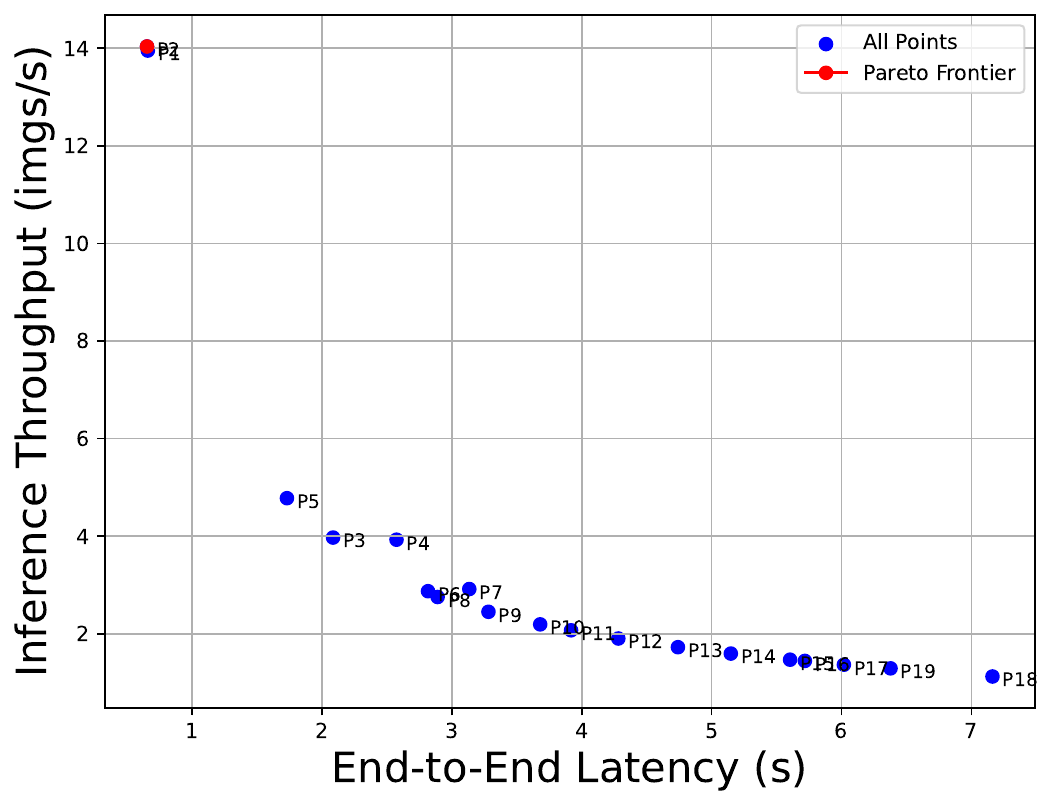}
    \caption{InceptionV3}
  \end{subfigure}
  \hfill
  \begin{subfigure}[b]{0.49\linewidth}
    \includegraphics[width=\linewidth]{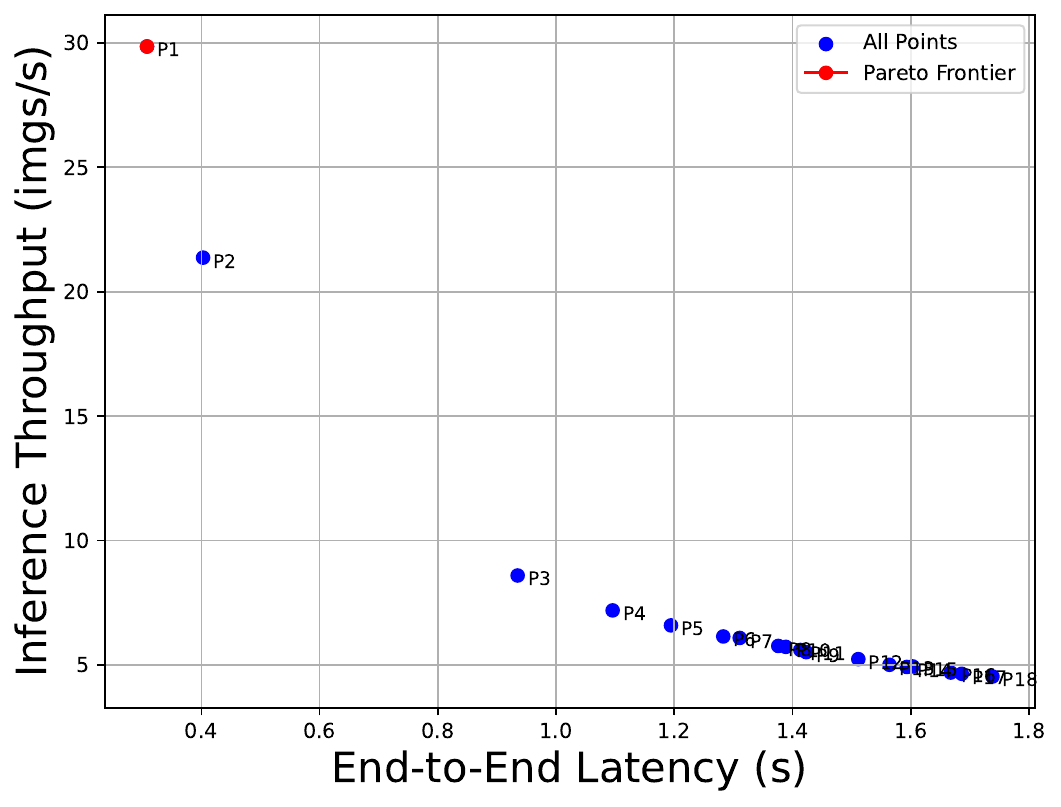}
    \caption{MobileNetV2}
  \end{subfigure}
  \begin{subfigure}[b]{0.49\linewidth}
    \includegraphics[width=\linewidth]{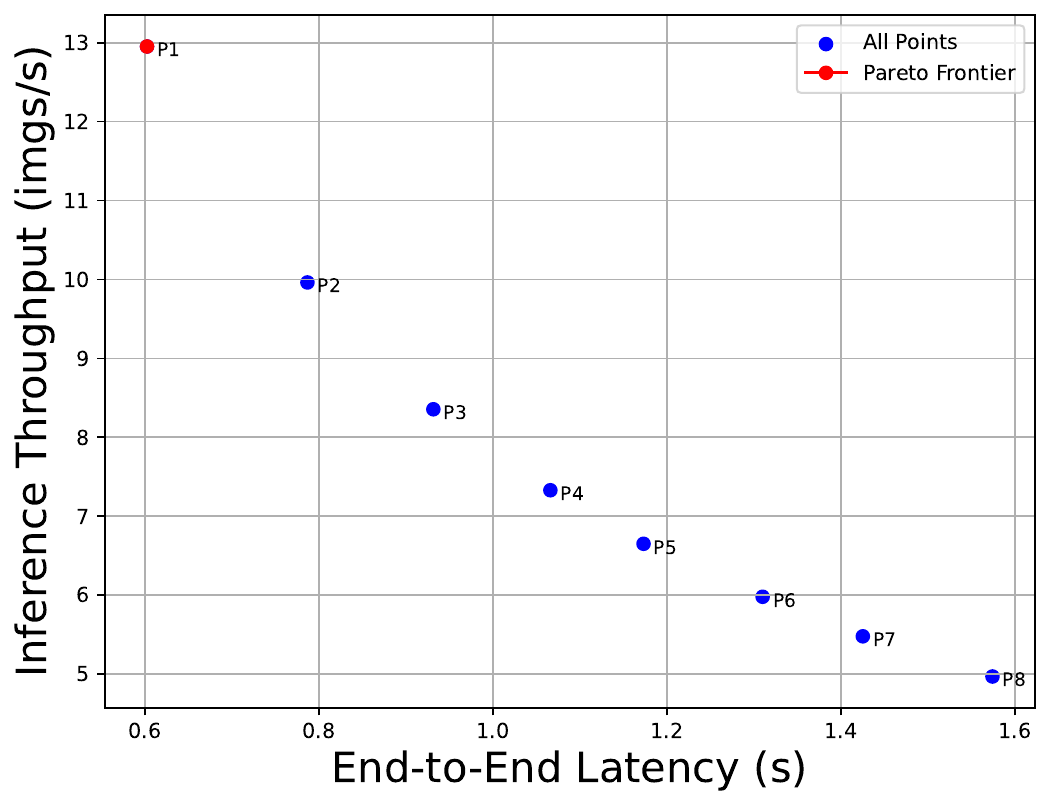}
    \caption{ResNet18}
  \end{subfigure}
  \hfill
  \begin{subfigure}[b]{0.49\linewidth}
    \includegraphics[width=\linewidth]{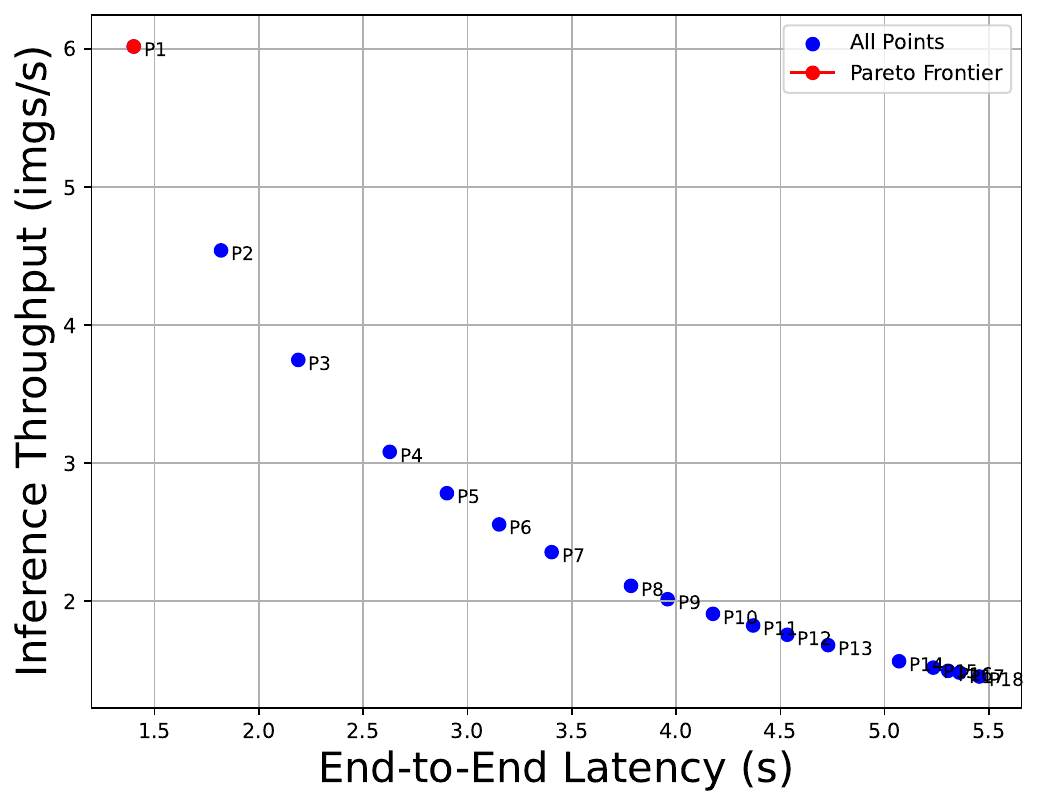}
    \caption{ResNet50}
  \end{subfigure}
  \hfill
  \begin{subfigure}[b]{0.49\linewidth}
    \includegraphics[width=\linewidth]{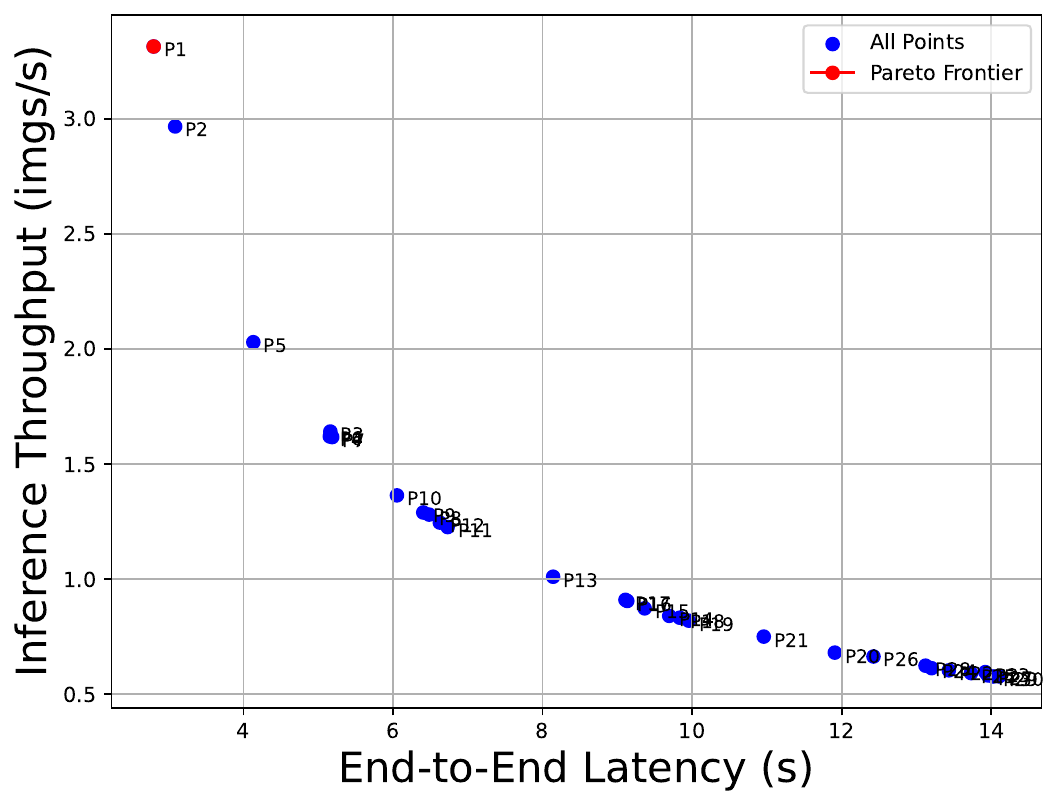}
    \caption{VGG16}
  \end{subfigure}
  \caption{Pareto Frontiers for DNN Partitioning on a Pi-to-GPU configuration.}
  \label{fig:pi-gpu}
\end{figure}

\section{Evaluation and Results}

All experiments presented in this paper utilize an inference batch size of eight; results for other batch sizes were omitted for brevity. The following analysis is based on ParetoPipe's lightweight communication backend (Sections~\ref{sec:pareto} and~\ref{sec:delay}), followed by a detailed comparison with PyTorch RPC in Section~\ref{sec:comparison}.

\subsection{Pareto Efficiency Front Analysis:}
\label{sec:pareto}
To systematically evaluate the trade-off between end-to-end inference latency (s) and inference throughput (imgs/s), we conducted a Pareto front analysis across six widely used DNNs listed in Table~\ref{tab:model_info}. The experiments were performed under two deployment scenarios: (1) Pi-to-Pi, where both parts of the model are executed on Raspberry Pi devices, and (2) Pi-to-GPU, where the first part runs on a Raspberry Pi and the second part is offloaded to a GPU-enabled edge server.
Figures \ref{fig:pi-pi} and \ref{fig:pi-gpu} illustrate the results for each model, where the x-axis represents end-to-end latency (seconds) and the y-axis denotes inference throughput (images per second). Each blue dot corresponds to a unique partition point, while the red line connects the Pareto-optimal points, which represent the best trade-offs where improvements in one metric cannot be achieved without sacrificing the other.


Lightweight models such as MobileNetV2 and AlexNet generally outperform heavier ones like VGG16 and ResNet50 in throughput and latency. In a Pi-to-Pi configuration (Figure \ref{fig:pi-pi}), optimal performance requires balancing the computational load. Models with evenly distributed layers like AlexNet, InceptionV3 and VGG16 are best split symmetrically, while MobileNetV2, ResNet18 and ResNet50 require an asymmetric split. For example, MobileNetV2 achieves maximum throughput when partitioned at block 3 (P3), offloading 18 out of 21 blocks to the second device, since its initial blocks are the most computationally intensive. These results highlight the significance of block-wise profiling discussed in Section~\ref{sec:profiling}.

By offloading computation to a GPU-enabled server, the Pi-to-GPU configuration (Figure \ref{fig:pi-gpu}) significantly boosts inference throughput and reduces latency, especially for larger models like ResNet50, VGG16, and InceptionV3. Since the workload is compute-bound, this aggressive offloading approach is the dominant optimization strategy, which results in a sparse Pareto front. For example, MobileNetV2 achieves the best performance when partitioned at block 1 (P1), offloading 20 out of 21 blocks to the GPU-enabled edge server.

\begin{figure}[tbp]
  \centering
  \begin{subfigure}[b]{0.49\linewidth}
    \includegraphics[width=\linewidth]{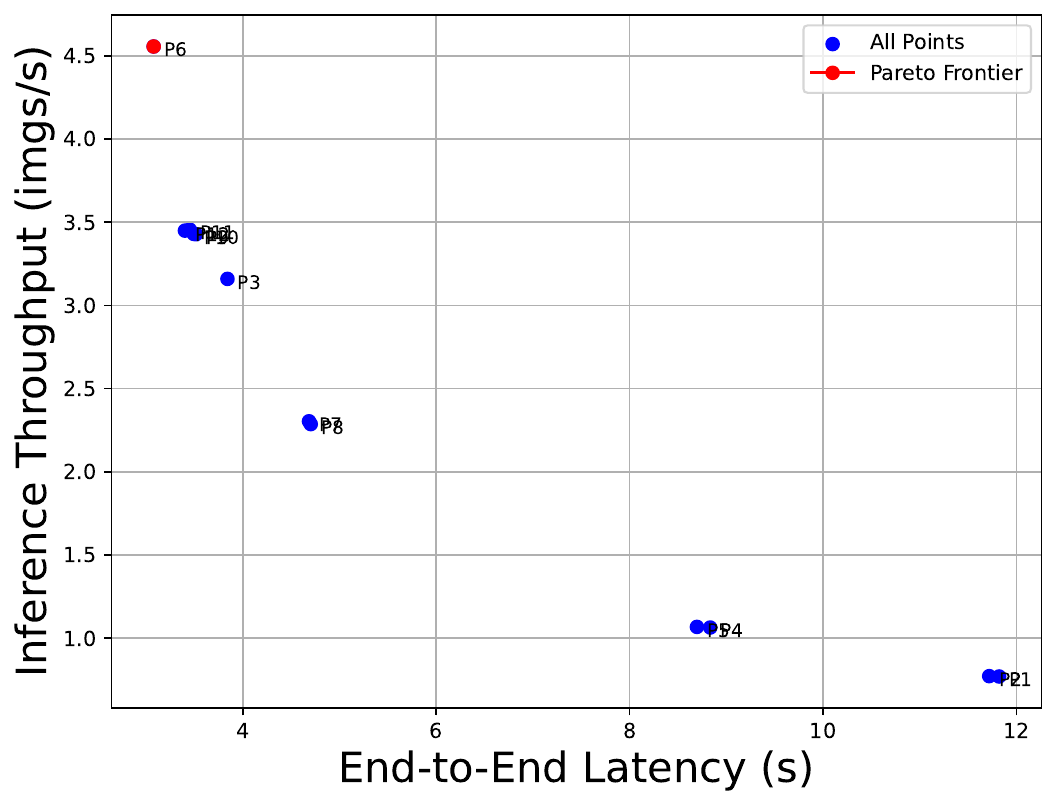}
    \caption{AlexNet}
  \end{subfigure}
  \begin{subfigure}[b]{0.49\linewidth}
    \includegraphics[width=\linewidth]{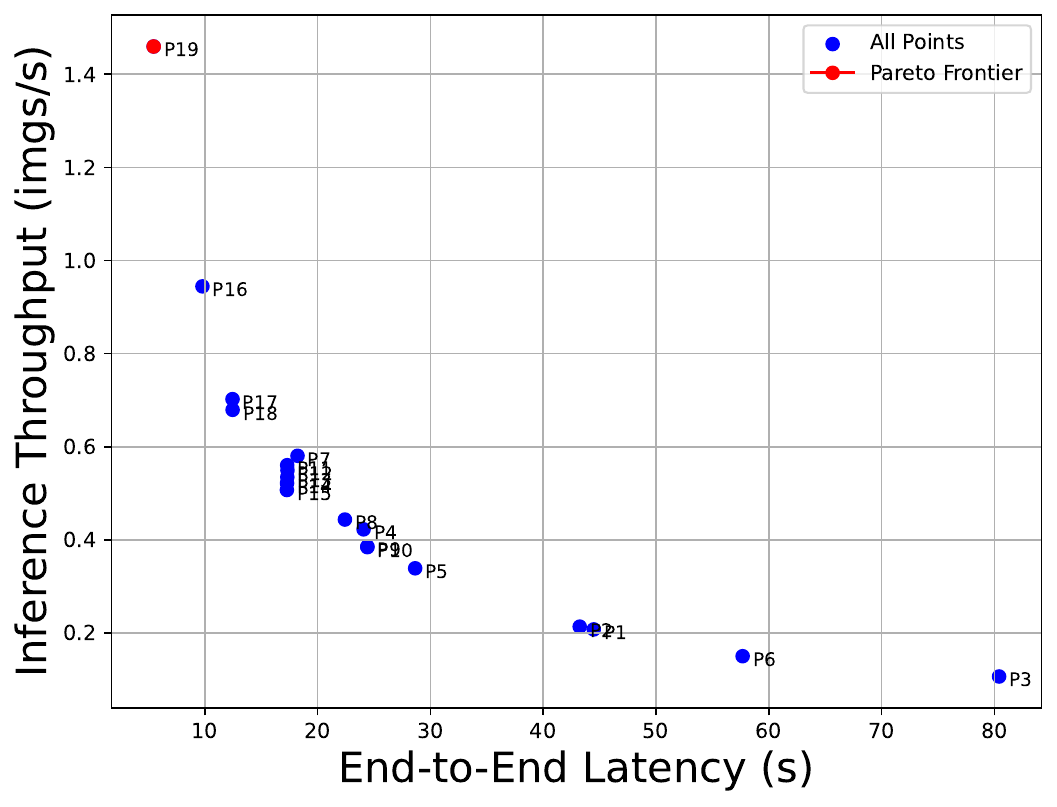}
    \caption{InceptionV3}
  \end{subfigure}
  \begin{subfigure}[b]{0.49\linewidth}
    \includegraphics[width=\linewidth]{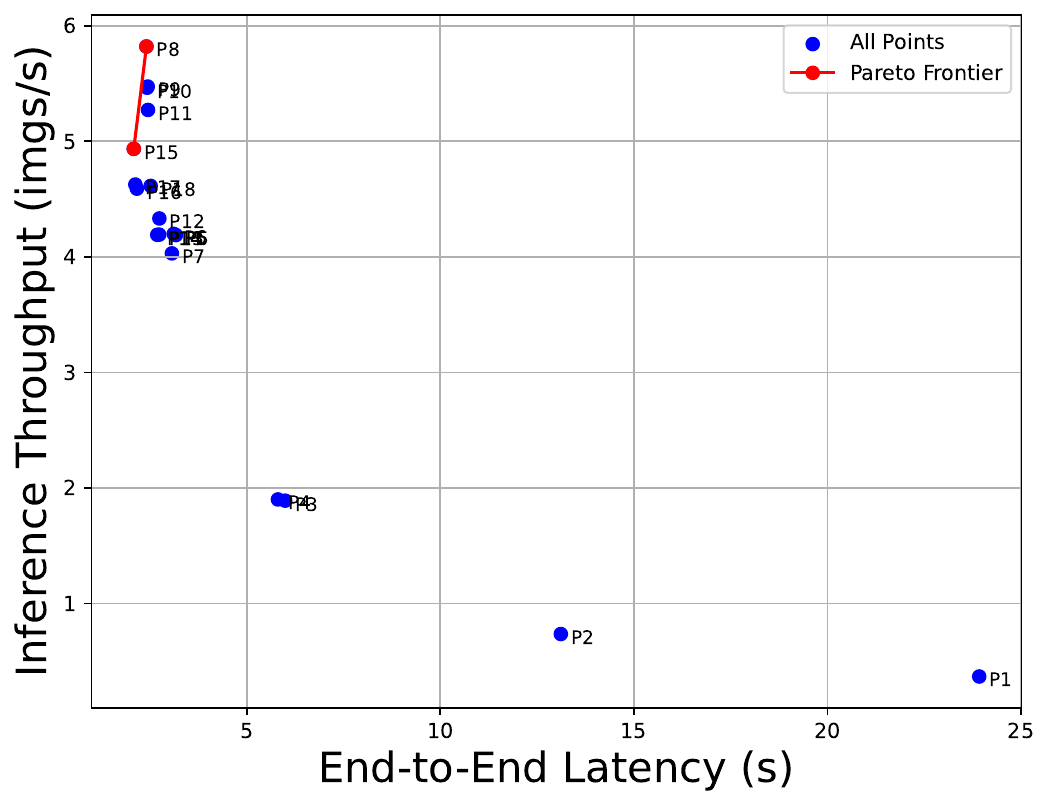}
    \caption{MobileNetV2}
  \end{subfigure}
  \begin{subfigure}[b]{0.49\linewidth}
    \includegraphics[width=\linewidth]{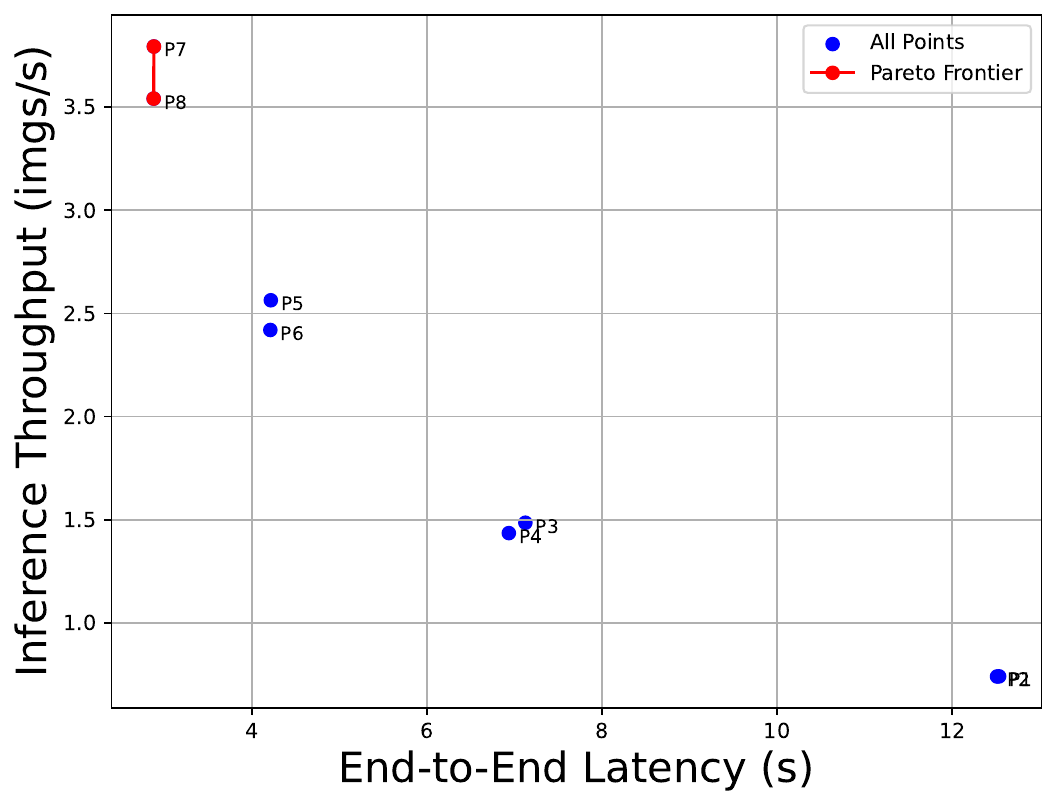}
    \caption{ResNet18}
  \end{subfigure}
  \begin{subfigure}[b]{0.49\linewidth}
    \includegraphics[width=\linewidth]{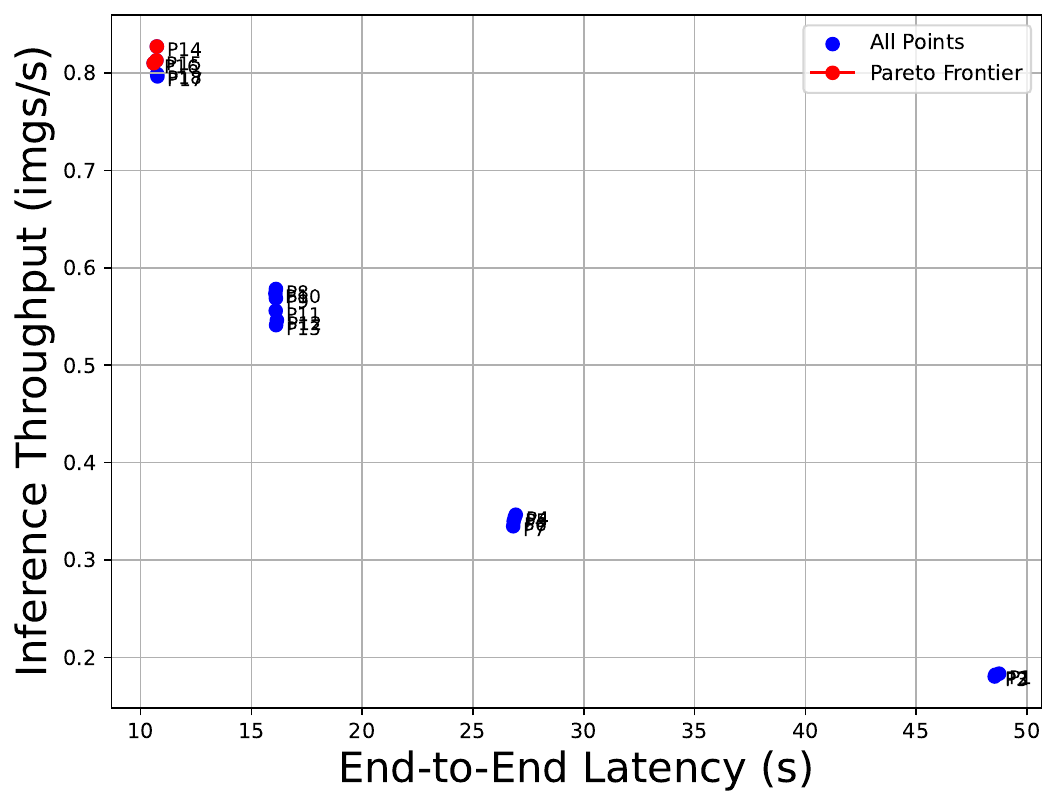}
    \caption{ResNet50}
  \end{subfigure}
  \begin{subfigure}[b]{0.49\linewidth}
    \includegraphics[width=\linewidth]{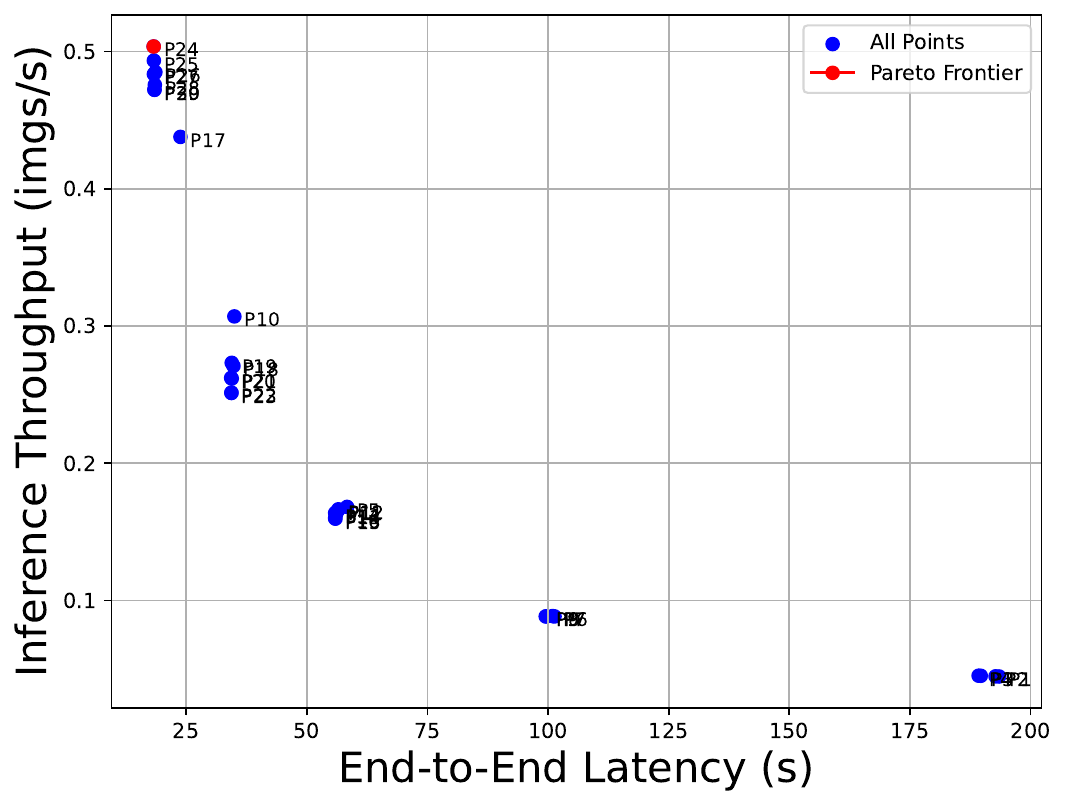}
    \caption{VGG16 }
  \end{subfigure}
  \caption{Pi-to-Pi Pareto Frontier under simulated network duress (200ms Latency, 5 Mbit/s Bandwidth).}
  \label{fig:pi-pi_delayNB}
\end{figure}

\begin{figure}[tbp]
  \centering
  \begin{subfigure}[b]{0.49\linewidth}
    \includegraphics[width=\linewidth]{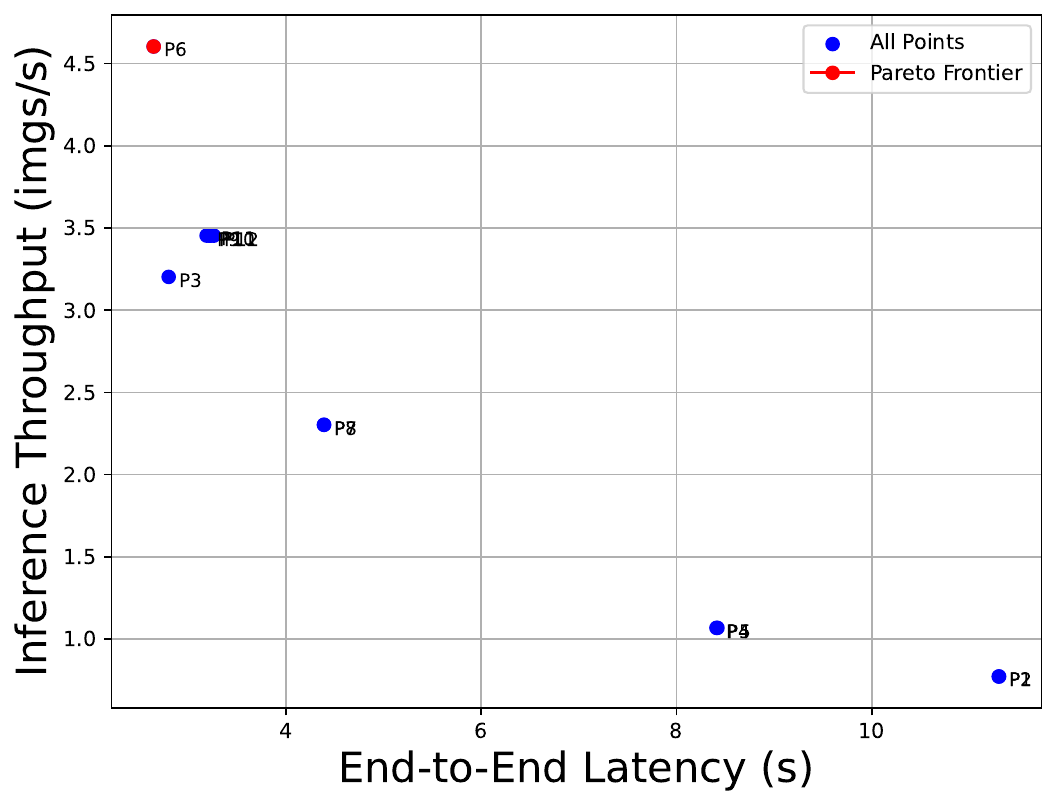}
    \caption{AlexNet}
  \end{subfigure}
  \hfill
  \begin{subfigure}[b]{0.49\linewidth}
    \includegraphics[width=\linewidth]{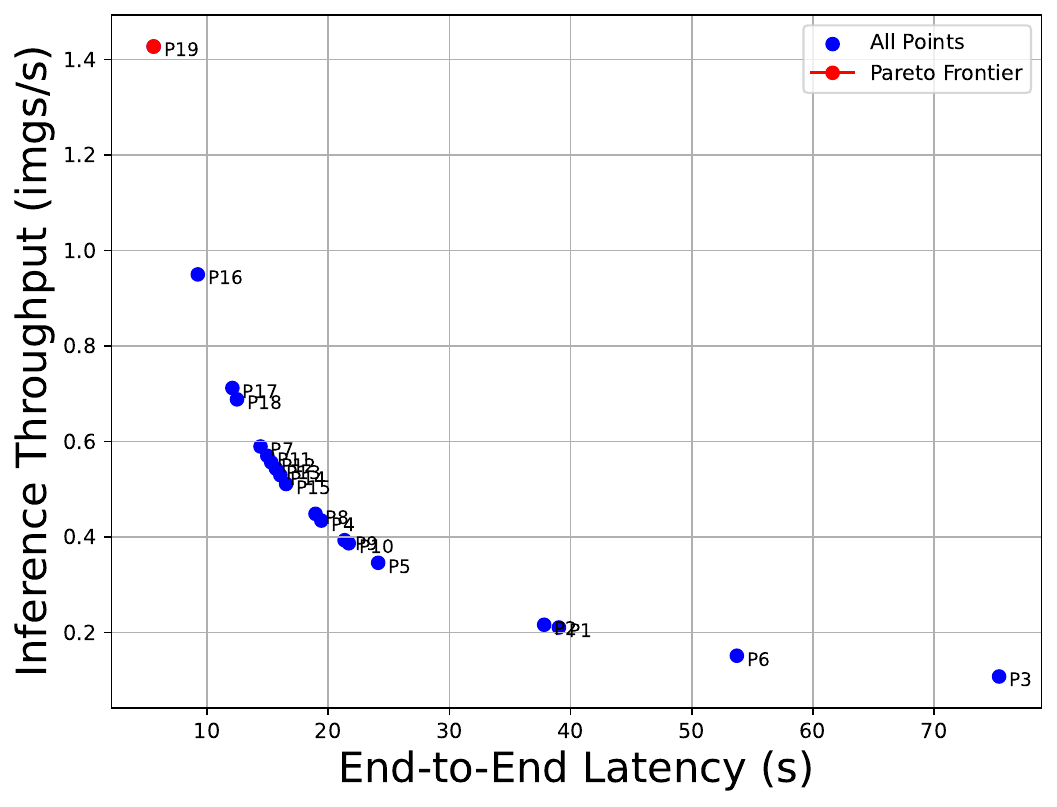}
    \caption{InceptionV3}
  \end{subfigure}
  \hfill
  \begin{subfigure}[b]{0.49\linewidth}
    \includegraphics[width=\linewidth]{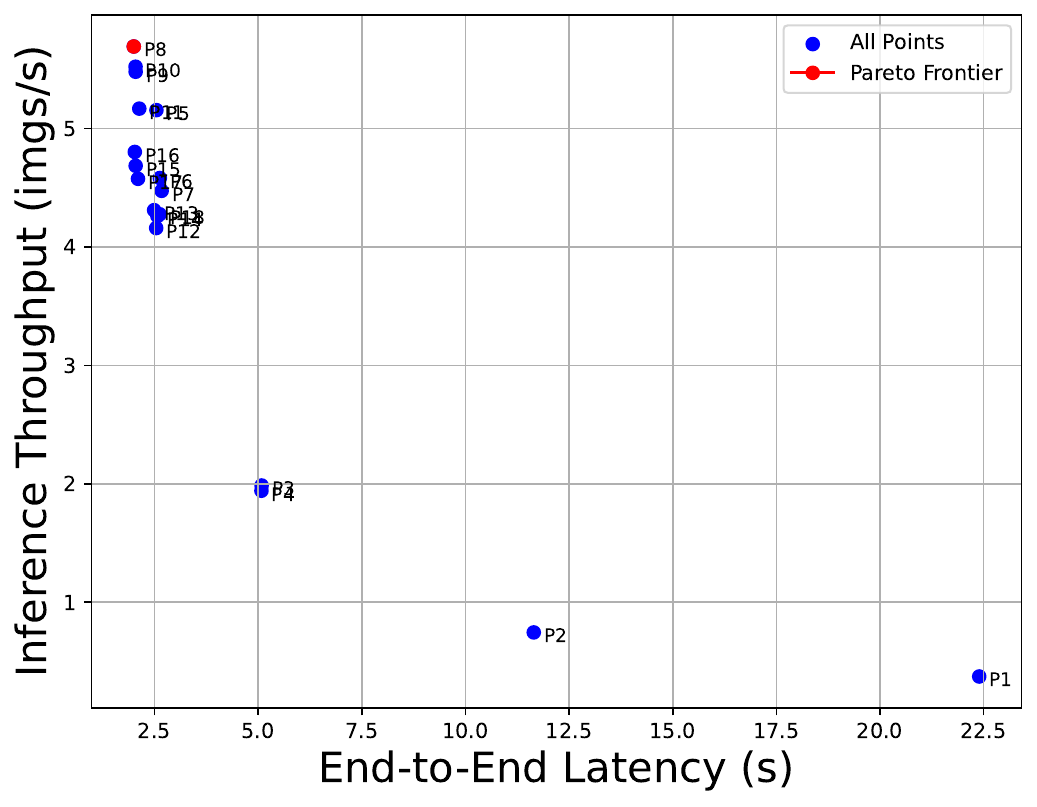}
    \caption{MobileNetV2}
  \end{subfigure}
  \begin{subfigure}[b]{0.49\linewidth}
    \includegraphics[width=\linewidth]{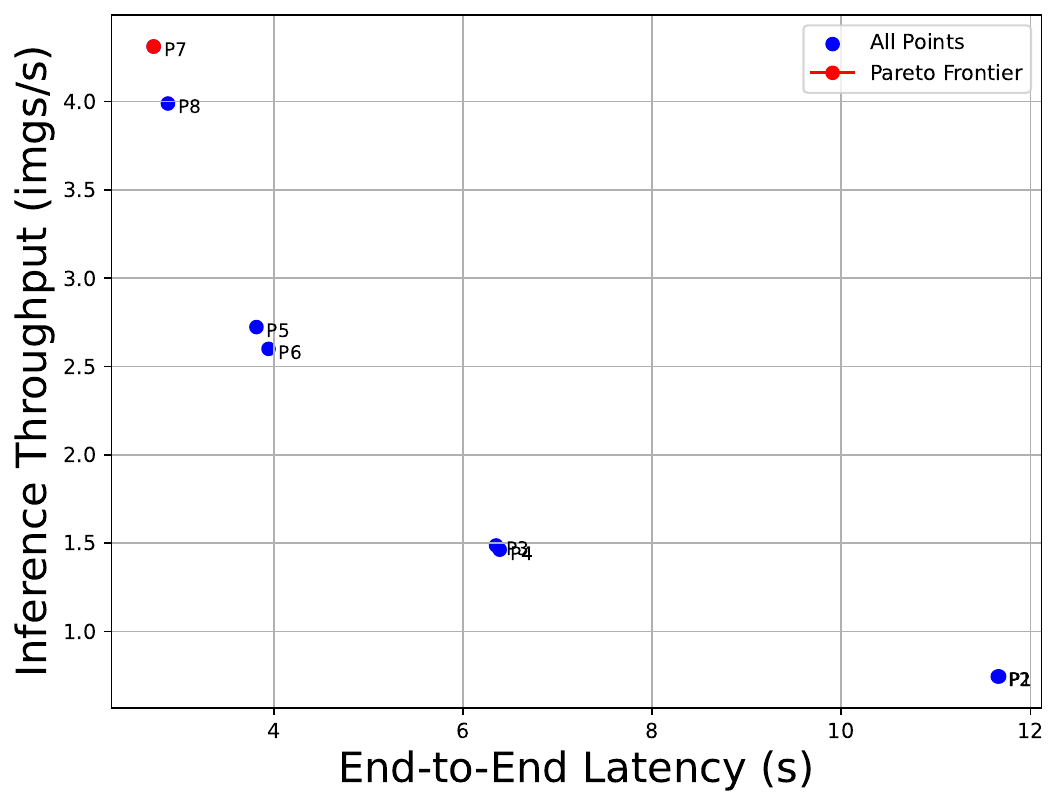}
    \caption{ResNet18}
  \end{subfigure}
  \hfill
  \begin{subfigure}[b]{0.49\linewidth}
    \includegraphics[width=\linewidth]{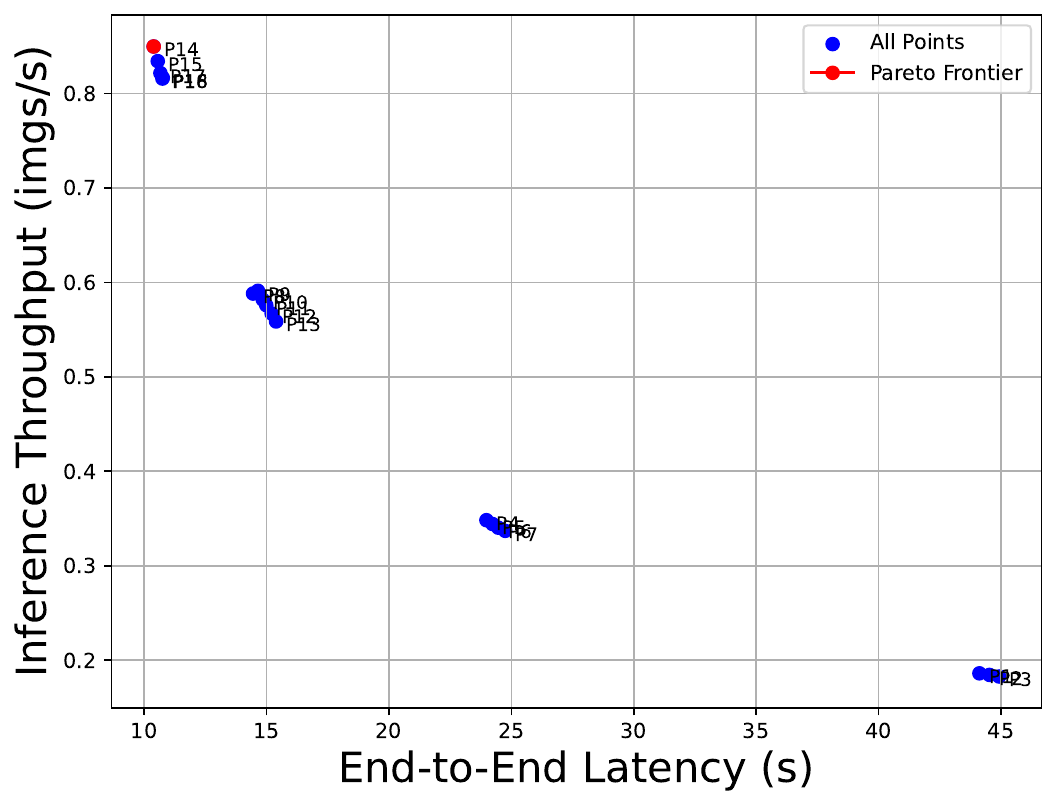}
    \caption{ResNet50}
  \end{subfigure}
  \hfill
  \begin{subfigure}[b]{0.49\linewidth}
    \includegraphics[width=\linewidth]{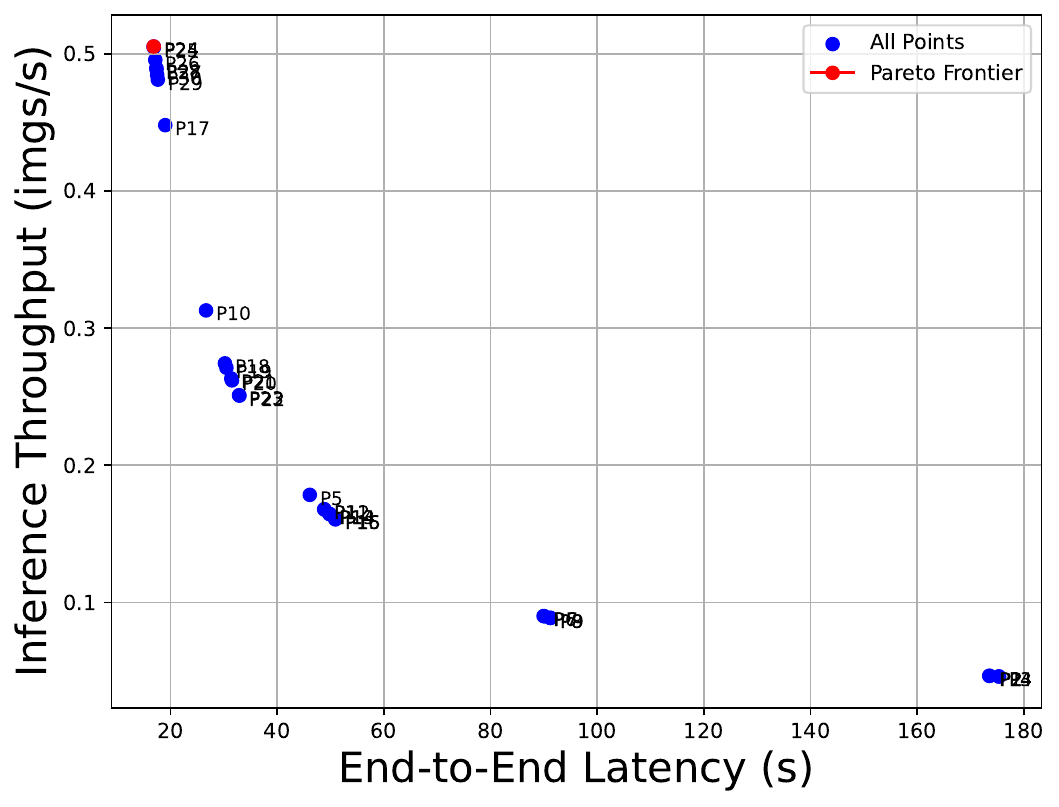}
    \caption{VGG16}
  \end{subfigure}
  \caption{Pi-to-GPU Pareto Frontier under simulated network duress (200ms Latency, 5 Mbit/s Bandwidth).}
  \label{fig:pi-gpu_delayNB}
\end{figure}


\subsection{Impact of Network Delay and Bandwidth:}
\label{sec:delay}
Using the Linux traffic control (tc) utility on the first Raspberry Pi in the pipeline, we simulated adverse network conditions by introducing a 200ms round-trip latency and throttling bandwidth to 5 Mbit/s. These parameters were chosen to mimic challenging, real-world deployment scenarios, such as operating over a congested Wi-Fi network. We then re-evaluated all six DNN models under these constraints to observe how their respective Pareto frontiers shifted, revealing the new performance trade-offs when the network becomes the primary bottleneck.

Figures \ref{fig:pi-pi_delayNB} (Pi-to-Pi) and \ref{fig:pi-gpu_delayNB} (Pi-to-GPU) show that the imposed 200ms delay and reduced network bandwidth degrade performance for all models. The substantial communication overhead of transferring intermediate tensors over the constrained network shifts the entire Pareto frontier to higher latencies and lower throughputs. In contrast to the baseline results shown in Section~\ref{sec:pareto}, both Pi-to-Pi and Pi-to-GPU configurations exhibit a sparse Pareto front when the network is the bottleneck. 

While the GPU can execute its portion of the model orders of magnitude faster than a Raspberry Pi, its computational power is rendered almost irrelevant by the network bottleneck as shown in Figure \ref{fig:pi-gpu_delayNB}. The time taken to transfer the intermediate tensors now constitutes the dominant portion of the end-to-end latency, as the powerful GPU server spends most of its time idle, waiting for the intermediate data to arrive from the Raspberry Pi. This disproportionately punishes partitioning strategies that offload work early (e.g., P1 in MobileNetV2), as this results in the largest amount of intermediate data to be transferred. Thus, the benefit of fast GPU computation is completely overshadowed by the communication cost.

These findings underscore a critical principle for edge AI deployments: network conditions are a first-class performance bottleneck. Relying on benchmarks from ideal laboratory conditions can lead to grossly optimistic performance estimates and poor partitioning choices. In high-latency or low-bandwidth environments, the optimal split point may shift towards performing more computation on the initial edge device to minimize the size of the data transferred over the network. This analysis validates the need for network-aware partitioning algorithms that can adapt to real-world conditions.

\subsection{Performance and Resource Usage Breakdown:}
\label{sec:comparison}
\begin{figure}[t]
    \centering

    \begin{subfigure}[t]{\linewidth}
        \centering
        \includegraphics[width=\textwidth]{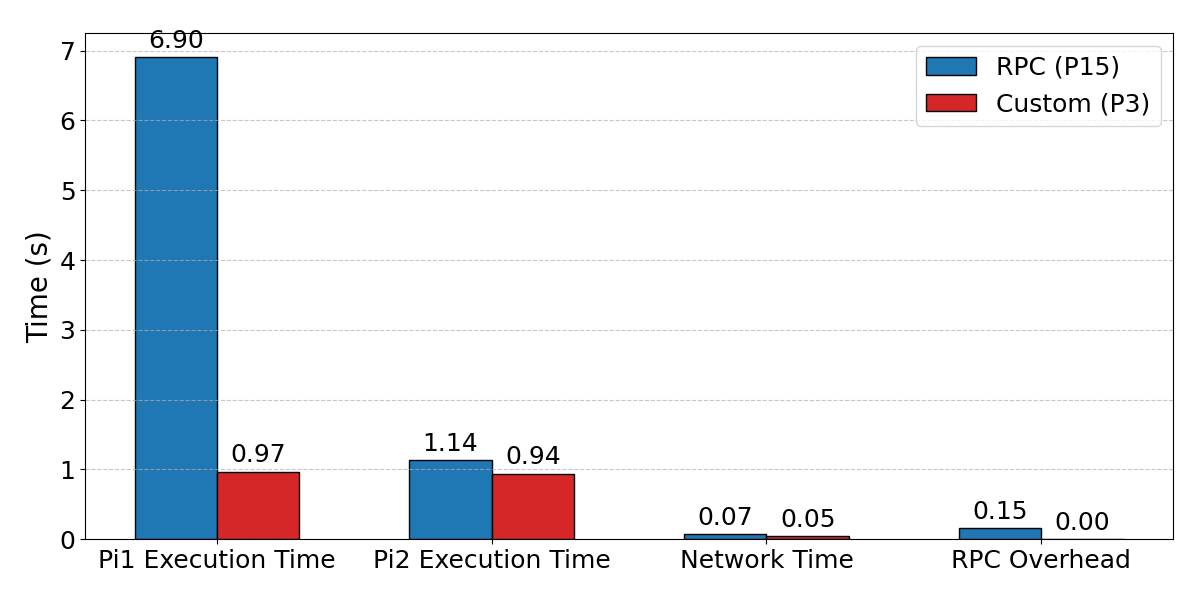}
        \caption{Execution Time Breakdown Comparison}
    \label{fig:rpc_vs_custom_exec_time_breakdown}
    \end{subfigure}
    
    \vspace{1em} 
    
    \begin{subfigure}[t]{\linewidth}
        \centering
        \includegraphics[width=\textwidth]{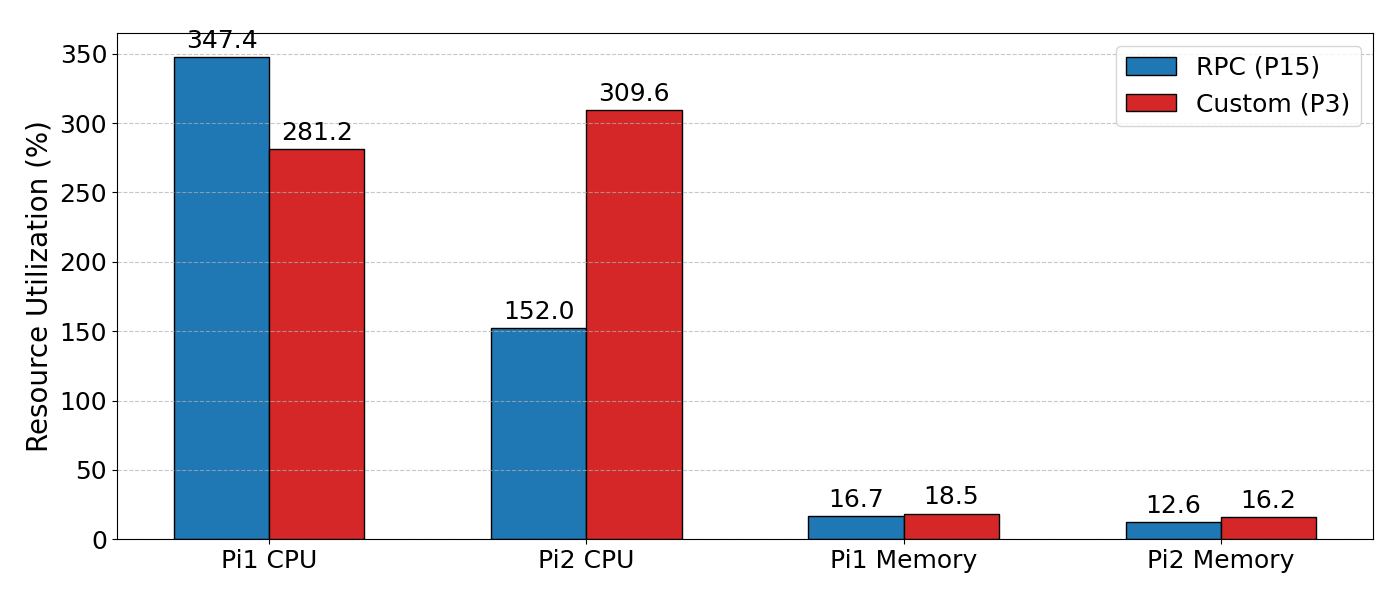}
        \caption{Resource Utilization Comparison}
    \label{fig:rpc_vs_custom_resource_util}
    \end{subfigure}

    \caption{Comparing PyTorch RPC vs. custom implementation at Pareto point with the highest throughput for MobileNetV2. For Pytorch RPC, partitioning at block 15 (P15) achieved an inference throughput of 5.1 images/s. For our custom implementation, partitioning at block 3 (P3) achieved a throughput of 7.8 images/s.} 
    \label{fig:rpc_vs_custom_mobilenetv2}
\end{figure}

In this section, we present a detailed breakdown of inference performance and system resource utilization. First, we compare the impact of communication backends on the performance of distributed inference. As a case study, we partitioned MobileNetV2 across two Raspberry Pis (Pi-to-Pi) and mapped out its Pareto frontier using both PyTorch RPC and our custom socket-based implementation. We used PyTorch 1.9+ with TensorPipe backend for RPC. For the custom implementation, we used direct TCP sockets with tensor serialization.  Figure~\ref{fig:rpc_vs_custom_mobilenetv2} compares the performance breakdown and resource utilization\footnote{CPU utilization exceeding 100\% indicates multi-core usage (4 core = 400\% max)} of these two cases using partitioning points that maximize the inference throughput. We observe that our custom implementation is significantly more efficient than PyTorch RPC in terms of per-device execution time. Furthemore, the custom backend eliminates RPC's coordination overhead, making it advantageous for edge inference. Overall, our custom implementation reduces the end-to-end inference latency by 76\% and improves the inference throughput by 53\% compared to PyTorch RPC.

Tables \ref{tab:custom_metrics} and \ref{tab:custom_metrics-(Pi-GPU)} show the performance breakdown and resource utilization of various DNN models using our socket-based communication backend in Pi-to-Pi and Pi-to-GPU settings, respectively. We highlight key partition points that lie on the Pareto frontier,  including examples that achieve the highest inference throughput or the lowest end-to-end latency. For example, in Table \ref{tab:custom_metrics}, AlexNet achieves the highest throughput at partitioning point P10 and the lowest end-to-end latency at partitioning point P11. We observe that both Raspberry Pi devices exhibit high CPU utilization for those partitioning points that lead to high throughput, across all models. On the other hand,  the partitioning points that achieve the lowest end-to-end inference latency are associated with imbalanced resource utilization between the two devices. In Table \ref{tab:custom_metrics-(Pi-GPU)}, we show only one entry per model since the Pareto frontier for the Pi-to-GPU setup consists of a single partitioning point. Together, these results provide the underlying empirical data that explain the latency-throughput trade-offs, while quantifying the resource costs for optimal partitioning strategies. 

\begin{table}[t]
\centering
\scriptsize
\setlength{\tabcolsep}{2pt}
\begin{tabular}{@{}p{2cm}p{.6cm}p{.6cm}p{.8cm}p{.8cm}p{.8cm}p{.8cm}p{.7cm}p{.7cm}@{}}
\toprule
\textbf{Model (Split)} & \textbf{Pi1-Exe(s)} & \textbf{Pi2-Exe(s)} & \textbf{Net-time(s)} & \textbf{Pi1-CPU Util(\%)} & \textbf{Pi2-CPU Util(\%)} & \textbf{Pi1 Mem(\%)}& \textbf{Pi2 Mem(\%)}\\
\midrule
AlexNet (P10)      & 0.451 & 0.427 & 0.050 & 272.7 & 271.9 & 24.84 & 21.76 \\
AlexNet (P11)      & 0.419 & 0.379 & 0.045 & 347.7 & 271.9 & 21.72 & 20.36 \\
InceptionV3 (P10)  & 2.873 & 2.766 & 0.055 & 328.8 & 322.7 & 22.24 & 20.76 \\
InceptionV3 (P19)  & 5.791 & 0.002 & 0.040 & 345.9 & 0.19   & 19.48 & 20.20 \\
MobileNetV2 (P3)   & 0.969 & 0.941 & 0.048 & 281.2 & 309.6 & 18.51 & 16.16 \\
MobileNetV2 (P17)  & 1.818 & 0.065 & 0.049 & 311.9 & 12.1  & 16.84 & 15.39 \\
ResNet18 (P2)      & 0.699 & 0.846 & 0.043 & 296.7 & 329.8 & 20.21 & 16.73 \\
ResNet18 (P6)      & 1.290 & 0.278 & 0.046 & 351.7 & 76.2  & 17.33 & 16.58 \\
ResNet50 (P5)      & 2.690 & 2.650 & 0.052 & 342.2 & 336.0 & 23.65 & 19.29 \\
ResNet50 (P15)     & 5.233 & 0.196 & 0.041 & 360.4 & 14.0  & 19.48 & 18.74 \\
VGG16 (P14)        & 6.827 & 6.319 & 0.056 & 328.9 & 305.7 & 35.43 & 32.87 \\
VGG16 (P29)        & 13.37 & 0.894 & 0.044 & 347.6 & 19.3 & 33.95 & 34.76 \\
\bottomrule
\end{tabular}
\caption{Performance breakdown with custom socket-based communication backend (Pi-to-Pi)}
\label{tab:custom_metrics}
\end{table}

\begin{table}[t]
\centering
\scriptsize
\setlength{\tabcolsep}{2pt}
\begin{tabular}{@{}p{1.8cm}p{.6cm}p{.6cm}p{.8cm}p{.8cm}p{.8cm}p{.8cm}p{.7cm}p{.7cm}@{}}
\toprule
\textbf{Model (Split)} & \textbf{Pi1 Exe(s)} & \textbf{GPU Exe(s)} & \textbf{Net-time(s)} &\textbf{Pi1-CPU Util(\%)} & \textbf{GPU Util(\%)} & \textbf{Pi1 Mem(\%)} & \textbf{GPU Mem(\%)} \\
\midrule
AlexNet (P3)      & 0.955 & 0.009 & 0.129 & 242.6  & 0.69 & 26.77  & 6.09 \\
InceptionV3 (P2)  & 0.538 & 0.021 & 0.200 & 311.3  & 1.89 & 28.5  & 4.09 \\
MobileNetV2(P1)  & 0.115 & 0.018 & 0.114 & 203.6   & 3.75  & 18.35 & 5.45\\
ResNet18 (P1)     & 0.514 & 0.011 & 0.082 & 327.3 & 0.07 & 19.55  & 5.32 \\
ResNet50 (P1)     & 0.975 & 0.012 & 0.094 & 281.3 & 0.69 & 25.14  & 5.82 \\
VGG16 (P1)        & 1.017 & 0.009 & 1.007 & 354.6  & 2.79 & 55.8  & 4.34 \\
\bottomrule
\end{tabular}
\caption{Performance breakdown with custom socket-based communication backend (Pi-to-GPU).}
\label{tab:custom_metrics-(Pi-GPU)}
\end{table}

\section{Conclusion and Future Work}
We designed and developed ParetoPipe, an open-source pipeline partitioning framework that systematically analyzes the trade-offs between inference latency and throughput. Our comprehensive analysis on a heterogeneous testbed reveals that while offloading to powerful hardware is effective under ideal conditions, the Pareto frontier shifts dramatically under realistic network constraints. In addition, we show that Remote Procedure Call (RPC) frameworks, often used in distributed systems, can introduce significant latency overhead. Due to these factors, strategies optimized in a lab can fail in real-world deployments. Furthermore, our work highlights the significance of block-level DNN profiling. Understanding the computational and memory costs of individual blocks or groups of layers is crucial for determining optimal partition points that balance the workload and minimize communication overhead. 

For future work, we plan to extend ParetoPipe to develop adaptive partitioning algorithms that can dynamically select an optimal split point at runtime in response to fluctuating network conditions. We also intend to incorporate energy efficiency as an additional objective, creating a more holistic view of performance. Finally, we will explore more complex, multi-device pipeline topologies and investigate hybrid partitioning methods suitable for emerging architectures like Transformers.

\section*{Acknowledgment} 

This material is based upon work partly supported by the UTSA College of AI, Cyber, and Computing's HEB Undergraduate Research Fellowship.

\end{document}